\documentclass[usenatbib, a4paper]{mn2e}

\usepackage{graphicx,fancyhdr,natbib,subfigure}
\usepackage{epsfig, epsf}
\usepackage{amsmath, cancel, amssymb}
\usepackage{lscape, longtable, caption}
\usepackage{dcolumn}
\usepackage{bm}
\usepackage{hyperref,ifthen}
\usepackage{verbatim}
\usepackage{color}
\usepackage[usenames,dvipsnames]{xcolor}

\def\jcap{\ref@jnl{J. Cosmology Astropart. Phys.}}%

\def \kms {{\rm ~km~s}^{-1}}

\def \gsim { \lower .75ex \hbox{$\sim$} \llap{\raise .27ex \hbox{$>$}} }
\def \lsim { \lower .75ex \hbox{$\sim$} \llap{\raise .27ex \hbox{$<$}} }

\newcommand{\simgt}{\raisebox{-0.6ex}{$\,\stackrel
        {\raisebox{-.2ex}{$\textstyle >$}}{\sim}\,$}}

\newcommand{\lya}{Ly$\alpha$\ }
\newcommand{\lyb}{Ly$\beta$\ }

\newcommand{\ciii}{C\,{\sc iii}]\ }
\newcommand{\civ}{C\,{\sc iv}\ }
\newcommand{\SiIV}{Si\,{\sc iv}\ }
\newcommand{\mgii}{Mg\,{\sc ii}\ }
\newcommand{\feii}{Fe\,{\sc ii}\ }

\newcommand{\halpha}{H\,$\alpha$\ }
\newcommand{\hbeta}{H\,$\beta$\ }
\newcommand{\hgamma}{H\,$\gamma$\ }
\newcommand{\hdelta}{H\,$\delta$\ }
\newcommand{\oi}{[O\,{\sc i}]\ }
\newcommand{\oii}{[O\,{\sc ii}]\ }
\newcommand{\oiii}{[O\,{\sc iii}]\ }

\newcommand{\nv}{N\,{\sc v}\ }
\newcommand{\nev}{Ne\,{\sc v}\ }
\newcommand{\neiii}{[Ne\,{\sc iii}]\ }
\newcommand{\aliii}{Al\,{\sc iii}\ }
\newcommand{\siiii}{Si\,{\sc iii}]\ }

\begin{document}

\title[Extremely Red Quasars: Optical Spectra]
{Extremely Red Quasars from SDSS, BOSS and WISE: \\
Classification of Optical Spectra}

\author[Ross et al.]
{Nicholas P. Ross$^{1,2}$\thanks{email: npross@physics.drexel.edu}, 
Fred Hamann$^3$, Nadia L. Zakamska$^{4}$, Gordon T. Richards$^{1,5}$, 
\newauthor Carolin Villforth$^{3,6}$, Michael A. Strauss$^7$, Jenny E. Greene$^{7,8}$, Rachael Alexandroff$^{4}$, 
\newauthor W. Niel Brandt$^{9,10}$, Guilin Liu$^{4}$, Adam D. Myers$^{11,5}$, Isabelle P\^{a}ris$^{12}$, 
\newauthor Donald P. Schneider$^{9,10}$\\
$^1$Department of Physics, Drexel University, 3141 Chestnut Street, Philadelphia, PA 19104, USA\\
$^2$Institute for Astronomy, SUPA\footnote{Scottish Universities Physics Alliance}, University of Edinburgh, Royal Observatory, Edinburgh, EH9 3HJ, United Kingdom\\
$^3$Department of Astronomy, University of Florida, Gainesville, FL 32611-2055, USA\\
$^4$Department of Physics \& Astronomy, Johns Hopkins University, 3400 N. Charles St., Baltimore, MD 21218, USA\\
$^5$Max-Planck-Institut f\"{u}r Astronomie, K\"{o}nigstuhl 17, D-69117 Heidelberg, Germany \\
$^6$SUPA, School of Physics and Astronomy, University of St Andrews, North Haugh, St Andrews, KY16 9SS, UK\\
$^7$Department of Astrophysical Sciences, Princeton University, Princeton, NJ 08544, USA\\
$^8$Alfred P. Sloan Fellow\\
$^{9}$Department of Astronomy and Astrophysics, The Pennsylvania State University, University Park, PA 16802, USA\\
$^{10}$Institute for Gravitation and the Cosmos, The Pennsylvania State University,   University Park, PA 16802, USA\\
$^{11}$Department of Physics and Astronomy, University of Wyoming, Laramie WY 82071, USA\\
$^{12}$INAF - Osservatorio Astronomico di Trieste, Via G. B. Tiepolo 11, I-34131 Trieste, Italy\\
}

\maketitle
\begin{abstract}
Quasars with extremely red infrared-to-optical colours are an interesting population that can test ideas about quasar evolution as well as orientation, obscuration and geometric effects in the so-called AGN unified model. To identify such a population we match the quasar catalogues of the Sloan Digital Sky Survey (SDSS), the Baryon Oscillation Spectroscopic Survey (BOSS) to the Wide-Field Infrared Survey Explorer (WISE) to identify quasars with extremely high infrared-to-optical ratios. We identify 65 objects with $r_{\rm AB}-W4_{\rm Vega}>14$ mag (i.e., $F_\nu({\rm 22\mu m})/F_\nu(r) \simgt 1000$). This sample spans a redshift range of $0.28 < z < 4.36$ and has a bimodal distribution, with peaks at $z\sim0.8$ and $z\sim2.5$. It includes three $z>2.6$ objects that are detected in the $W4$-band but not $W1$ or $W2$ (i.e., ``W1W2-dropouts''). The SDSS/BOSS spectra show that the majority of the objects are reddened Type 1 quasars, Type 2 quasars (both at low and high redshift) or objects with deep low-ionization broad absorption lines (BALs) that suppress the observed $r$-band flux. In addition, we identify a class of Type 1 permitted broad-emission line objects at $z\simeq 2-3$ which are characterized by emission line rest-frame equivalent widths (REWs) of $\gtrsim$150\AA , much larger than those of typical quasars. In particular, 55\% (45\%) of the non-BAL Type 1s with measurable CIV in our sample have REW(CIV) $>$ 100 (150)\AA , compared to only 5.8\% (1.3\%) for non-BAL quasars in BOSS. These objects often also have unusual line ratios, such as very high N\,{\sc v}/Ly$\alpha$ ratios. These large REWs might be caused by suppressed continuum emission analogous to Type 2 quasars; however, there is no obvious mechanism in standard Unified Models to suppress the continuum without also obscuring the broad emission lines.
\end{abstract}


\section{Introduction}

Quasars are the most luminous non-transient objects in the Universe,
with $L_{\rm bol}$ reaching $\sim$$10^{47-48}$ erg s$^{-1}$, powered
by accretion of matter onto supermassive black holes.  To explain how
quasars are fueled and become active, models
\citep[e.g.,][]{Sanders88, Hopkins05b, Hopkins06a} suggest that
mergers between gas-rich galaxies produce starbursts and drive gas to
the inner galactic nuclear regions. This activity fuels the growth of
the supermassive black holes, which actively accrete matter at this
stage, with much of the black hole growth happening in this
ultra-violet (UV) and optically obscured phase. Feedback, potentially
from a radiation-driven quasar wind, leads to removal of gas and dust
from the central region, allowing the quasar to be seen in the UV and
optical. These major-merger models are invoked to explain the
obscuration/reddening observations noted in e.g., \citet{Urrutia08},
\citet{Glikman12}, and predict that red/obscured quasars should appear
preferentially during an early stage of quasar evolution. 
However, the exact nature of the AGN triggering mechanism is not 
straightforward, as observations find that levels of disturbance in
AGN hosts are consistent with those of inactive control samples
\citep[e.g.][]{Kocevski12, Villforth14}, or only slightly elevated at
low-$z$ \citep{Rosario15}.

Quasars appear with a wide range of observational
properties. Intrinsically luminous quasars may or may not appear
bright at optical, ultra-violet and X-ray wavelengths. Much of this
variance can be explained in the context of the geometry-based
`unification models' \citep{Antonucci93}. If gas and dust are present
near the active nucleus but do not completely surround it, then some
lines of sight to the nucleus are clear, whereas others are
blocked. In the former case, the observer can directly view the
emission from the accretion disk, with the quasar appearing bright at
X-ray, ultra-violet and optical wavelengths; these objects are termed
unobscured active galactic nuclei. Often these objects display broad
(several thousand $\kms$) emission lines in the optical spectra and
are classified as `Type 1' quasars \citep{Khachikian_Weedman74}.  Less
than complete blockage of the quasar continuum by dust will redden the
spectrum \citep[e.g.,][]{Richards03, Krawczyk15}.

This contrasts with the situation where the observer's line of sight
is blocked by circumnuclear clouds; in this case X-rays are absorbed
by the intervening gas and ultra-violet and optical photons are
scattered and absorbed by the intervening dust. At optical
wavelengths, often the only signature of nuclear activity in this case
is strong, narrow emission lines produced in the material illuminated
by the quasar along unobscured directions. These objects with weak
ultra-violet and optical continua are obscured AGN, and if narrow
UV/optical emission lines are present, are designated Type 2 sources
\citep{Antonucci_Miller85, Smith02, Zakamska03, Brandt_Hasinger05,
Reyes08}. Thus, obscuration, as part of an evolutionary phase or an
orientation effect, is critical in determining the observed properties
of quasars. However, because of their faintness at rest-frame optical
and ultra-violet wavelengths, identifying obscured objects, at the
peak of quasar activity $z\sim 2$, remains challenging \citep{Stern02,
Norman02, Alexandroff13, Greene14}.

The circumnuclear clouds of gas and dust absorb X-ray, ultra-violet
and optical radiation from the quasar and re-emit this energy
thermally at infrared wavelengths. This is why unobscured quasars have
similar luminosities at near- and mid-infrared wavelengths
($\sim2-30\mu$m in the rest-frame) as they do in the optical and in
the ultra-violet \citep[][]{Elvis94, Richards06b, Polletta08,
Elvis10}. Extremely obscured and dusty objects, in which optical
emission is partially extincted or completely blocked, are therefore
expected to show much higher infrared-to-optical ratios than
unobscured quasars. Indeed, a host of previous studies have used a
{\it K}-band excess selection \citep{Chiu07, Maddox08, Jurek08,
Nakos09, Souchay09, Wu10, Peth11, Wu11,Maddox12, Fynbo13, Wu13}, a
near-infrared+radio selection \citep{Glikman04, Glikman07, Glikman12,
Glikman13} or a mid-infrared selection \citep{Lacy04, Stern05,
Martinez-Sansigre06, Richards09b, Donley12, Stern12, Banerji13,
Assef13} to identify obscured (as well as unobscured) AGN.

In this paper, we describe a search for quasars with extremely red
colours in order to study their range of spectral properties.  We use
optical photometry and spectroscopy from the Sloan Digital Sky Survey
\citep[SDSS;][]{York00} and the SDSS-III \citep{Eisenstein11} Baryon
Oscillation Spectroscopic Survey \citep[BOSS;][]{Dawson13} as well as
mid-infrared photometry from the Wide-Field Infrared Survey Explorer
\citep[WISE; ][]{Wright10}.  A companion paper (Hamann et al. 2015, in
prep.) presents a detailed investigation into a new class of objects,
the extreme rest-frame equivalent width quasars, which are introduced
here.

This paper is organized as follows. In Section~\ref{sec:data}, we
describe our datasets and sample selection, and in
Section~\ref{sec:key_props} we present the basis sample properties of
the extremely red quasars. In Section~\ref{sec:spectral_egs} we
discuss the optical spectra of the extremely red quasars, classifying
objects along the traditional lines of broad-line Type 1s, narrow-line
Type 2s and those with interesting absorption features. In
Section~\ref{sec:erews}, we introduce the new class of extreme
equivalent width quasars.  In Section~\ref{sec:spectral_selection} we
discuss the selection of the extremely red quasars and place these
objects in a broader physical and evolutionary context.  We conclude
in Section~\ref{sec:conclusions}.


\begin{figure}
  \includegraphics[width=8.20cm, height=8.20cm, trim=0.4cm 0.4cm 0.8cm 0.4cm, clip]
  {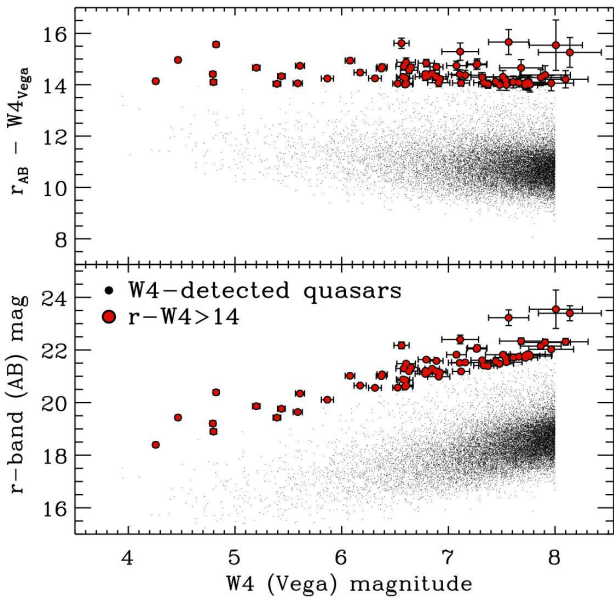}
  \centering
  \vspace{-10pt}
  \caption[]{{\it Top:} $r-W4$ colour of the $W4$-detected SDSS/BOSS quasars
    (black dots) as a function of $W4$ (i.e., 22$\mu$m) magnitude. The
    quasars with $r-W4>14$ are shown as large red circles. {\it Bottom:}
    The $r_{\rm AB}$-band magnitude versus $W4$. As a guide, $W4_{\rm
      Vega}=8.0$ mag corresponds to $W4_{\rm AB}=14.6$, or a flux density of
    $F_{\nu}\approx6$mJy \citep{Wright10}.}
  \label{fig:W4_vs_rband}
\end{figure}
\section{Data and Sample Selection}\label{sec:data}

    \subsection{Parent dataset}
    Our starting point is the spectroscopic quasar catalogues of the
    SDSS Seventh Data Release \citep[DR7Q;][]{Schneider10, Shen11} and the
    SDSS-III Tenth Data Release \citep[DR10Q;][]{Paris14}. SDSS quasar
    targets with $i\leq19.1$ were selected if the colours were consistent
    with being at redshift $z\lesssim3$, and $i\leq20.2$ objects were
    selected if $z\gtrsim$3, as outlined in \citet{Richards02}. BOSS
    quasar targets are selected to a magnitude limit of $g \leq 22.0$ or
    $r\leq 21.85$, with the primary goal to select quasars in the redshift
    range $2.2\leq z \leq 3.5$ as described by \citet[][and references
    therein]{Ross12}. In both SDSS and BOSS, quasar targets are also
    selected if they are matched within 2$''$ (1$''$ in the case of BOSS)
    of an object in the Faint Images of the Radio Sky at Twenty-cm (FIRST)
    catalogue of radio sources \citep{Becker95}. Both the SDSS DR7Q and
    BOSS DR10Q include quasars that were selected by algorithms other than
    the main quasar selections; these sources appear in the catalogue due
    to being targeted by the respective galaxy selections, being a
    `serendipitous' \citep{Stoughton02} or `special' \citep{DR4} target in
    SDSS, or an `ancillary' target in BOSS \citep[][]{Dawson13, DR12}.
    
    Luminous distant quasars are expected to be unresolved in optical
    ground-based observations, so we use point-spread function (PSF)
    magnitudes reported in the SDSS DR7 and BOSS DR10 quasar catalogues,
    corrected for Galactic extinction \citep{Schlegel98}. Using the 2.5m
    Sloan telescope \citep{Gunn06}, imaging data \citep{Gunn98} and
    spectra were obtained with the double-armed SDSS/BOSS spectrographs
    which have $R= \lambda$/FWHM$\sim$2000 \citep{Smee13}. Redshifts are
    measured from the spectra using the methods described in
    \citet{Bolton12} and the data products are detailed in the SDSS DR7
    \citep[][]{DR7} and DR10 \citep{DR10} papers.
    
    The SDSS DR7 and BOSS DR10 quasar catalogues include \hbox{105,783}
    objects across 9380 deg$^{2}$, and \hbox{166,583} objects across
    6,370~deg$^{2}$, respectively, with \hbox{16,420} objects in common to
    both catalogues. Thus our superset of optical data has \hbox{255,946}
    objects, and we use the BOSS spectrum (which has higher S/N and a
    larger wavelength coverage) when there are duplicate spectra between
    the two catalogues.
    
    WISE mapped the sky in four filters centered at 3.4, 4.6, 12, and
    22$\mu$m ($W1$, $W2$, $W3$, and $W4$ bands), achieving $5\sigma$
    point-source sensitivities better than 0.08, 0.11, 1, and 6 mJy,
    respectively. The WISE Explanatory
    Supplement\footnote{\href{http://wise2.ipac.caltech.edu/docs/release/allwise/expsup/index.html}{wise2.ipac.caltech.edu/docs/release/allwise/expsup/index.html}}
    provides further details about the astrometry and photometry in the
    source catalogue. We retrieved photometric quantities for each of the
    four WISE bands. Hereafter, we use the notation SNR$_{{\rm W}x}$ for
    the signal-to-noise ratio and $Wx$ for the Vega-based WISE magnitudes,
    in band $x=1,2,3$ and 4.
    
    Because of established conventions, we report Sloan Digital Sky Survey
    (SDSS) $ugriz$ magnitudes on the AB zero-point system \citep{Oke83,
      Fukugita96}, while the Wide-Field Infrared Survey Explorer (WISE)
    $W1-4$ magnitudes are calibrated on the Vega system
    \citep{Wright10}. For WISE bands, $m_{\rm AB} = m_{\rm Vega} + \Delta
    m$ where $\Delta m=(2.699, 3.339, 5.174, 6.620)$ for $W1$, $W2$, $W3$
    and $W4$, respectively \citep{Cutri11}. We make use of the Explanatory
    Supplement to the WISE All-Sky Data Release, as well as the WISE
    AllWISE Data Release Products online.
    
    The WISE team has released two all sky catalogues: the 9 month cryogenic
    phase of the mission led to the WISE All-Sky (``AllSky'') Data
    Release; while the WISE AllWISE Data Release (``AllWISE'') combines
    the AllSky data with the NEOWISE program \citep{Mainzer11}. The
    resulting AllWISE dataset is deeper in the two shorter WISE-bands
    ($5\sigma$ point-source sensitivities now 0.054 and 0.071 mJy), and
    the data processing algorithms were improved in all four $W$-bands.
    Our analysis began with AllSky, moved to ALLWISE when it became
    available, but we use both since we had already performed visual
    inspections using AllSky.
   
    We match the sample of the \hbox{255,946} unique quasars in the
    SDSS DR7Q and BOSS DR10Q, and use both the AllSky and the AllWISE data
    available at the NASA/IPAC Infrared Science
    Archive (IRSA)\footnote{\href{http://irsa.ipac.caltech.edu/}{http://irsa.ipac.caltech.edu/}},
    with a matching radius of 2$''$. For the combined DR7Q+DR10Q, we find
    \hbox{203,680} matches; \hbox{102,083} objects (96.5\%) of the DR7Q
    and \hbox{111,779} (67\%) of the DR10Q are matched in one or more WISE
    bands. The difference in the percentage of matches is expected since
    BOSS is fainter than SDSS. Matching with DR10Q positions offset in
    Right Ascension and Declination by 6$''$, 12$''$ and 18$''$
    ($\sim$1, 2 and 3 times the WISE angular resolution in W1/2/3) return,
    respectively, 508, 893 and 873 matches within 2$''$, suggesting our
    false-positive matching rate is $\lesssim$1\% \citep[see
    also][]{Krawczyk13}.
    
    With a matching radius of 2$"$, \hbox{15,843} of the SDSS DR7 quasars,
    and \hbox{2,979} of the (optically fainter) BOSS DR10 quasars have a
    good match in the WISE $W4$ band, i.e., SNR$_{\rm W4} \geq3$, W4$<8.00$
    (close to the nominal 5$\sigma$ point source sensitivity of $W4=7.9$ mags,
    \citealt{Wright10}) and the WISE contamination and confusion flag,
    {\tt cc\_flags} set to ``0000'', suggesting the source is unaffected
    by known artifacts in all four bands. Table~\ref{tab:ERQ_key_numbers}
    presents an overview of the numbers of SDSS/BOSS quasars in the
    WISE-matched dataset.

    \begin{table}
      \begin{center}
        \begin{tabular}{lrrr}
          \hline
          \hline
          Description & Unique          & from  & from\\
          & objects          & SDSS  & BOSS\\
          \hline  
          DR7Q and DR10Q quasars$^{a}$                                            & \hbox{255 946}  & \hbox{105 783} & \hbox{166 583}\\
          $\quad \shortparallel$  + matched to WISE                       & \hbox{203 680}  & \hbox{102 083} & \hbox{111 779}\\
          $\quad \shortparallel \quad$ + SNR$_{\rm W4} \geq 3$       & \hbox{ 52 873}   & \hbox{ 41 922}  & \hbox{  10 951}\\
          $\quad \shortparallel \qquad$ + $W4-\sigma_{W4} \leq 8.00$    & \hbox{ 28 513}   & \hbox{ 24 041}  & \hbox{  4 472}\\
          $\quad \shortparallel \quad \qquad$ + $W4 \leq 8.00$          & \hbox{ 18 615}   & \hbox{ 16 035}  & \hbox{  2 580}\\
          $\quad \shortparallel \qquad \qquad$ + {\tt cc\_flags}     & \hbox{ $^{b}$17 744}  & \hbox{ 15 300}  & \hbox{  2 444}\\
          $\qquad \qquad \qquad$ and $r-W4 >13.0$     &                  286   &                  104   &               182 \\
          $\qquad \qquad \qquad$ and $r-W4 >13.5$     &                  136  &                   34    &                 102 \\
          $\qquad \qquad \qquad$ and $r-W4 >14.0$     &            {\bf 65}  &            {\bf 19}  &           {\bf 48} \\
          \hline
          \hline
        \end{tabular}
        \caption{Overview of the numbers of SDSS/BOSS quasars in the WISE-matched
          dataset. The third row gives the number of quasars which satisfy 
            $W4$ SNR $\geq 3$; the fourth row is the number of quasars 
            having a magnitude of $W4-\sigma_{W4} \leq 8.00$, while the 5th row
            is the number having $W4 \leq 8.00$.
          Numbers of objects passing the $r_{\rm AB}-W4_{\rm Vega}$ colour cut is
          given. $^{a}$There are objects common to both SDSS DR7Q and BOSS
          DR10Q that have WISE matches. Hence, the sum of the ``From SDSS'' and
          ``From BOSS'' columns is greater than the given total. $^{b}$``Good''
          $W4$ detections have SNR$_{\rm W4} \geq 3$ and $W4<8.00$ and 
          {\tt  cc\_flags} equal to ``0000''.}
        \label{tab:ERQ_key_numbers}
      \end{center}
      \vspace{-8pt}
    \end{table}

    \begin{figure}
      \includegraphics[width=8.2cm, trim=18mm 105mm 105mm 10mm, clip]
      {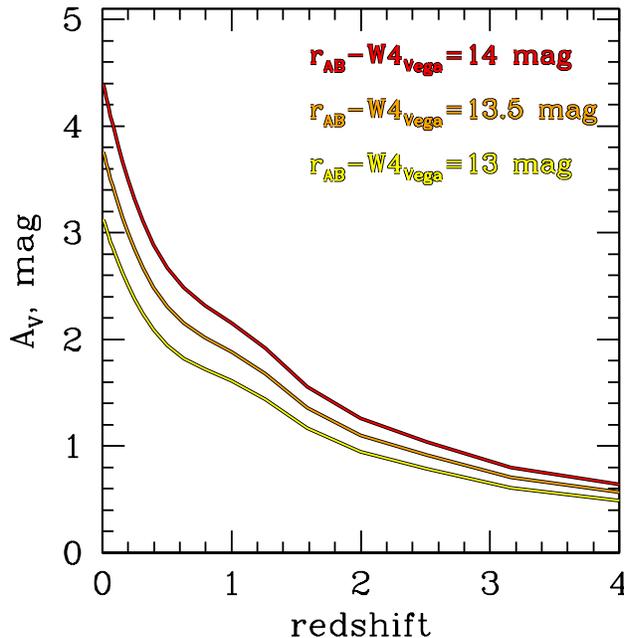}
      \centering
      \vspace{-8pt}
      \caption[]{
        The minimal amount of dust extinction required to produce high
        infrared-to-optical ratios consistent with our selection criteria, as
        a function of the redshift of the quasar. The \citet{Richards06b}
        spectral energy distribution for bright quasars is assumed to
        represent an unextincted mean quasar. We then apply varying amounts of
        SMC-like extinction \citep[from ][]{Weingartner01} and convolve with
        the respective filter curves. We assume $R_V \equiv
        \frac{A_V}{E(B-V)}=2.87$ \citep{Gordon98}. The three curves
        show the $A_V$ values required to generate colours of $r-W4>13$
        (yellow) $r-W4>13.5$ (orange) and $r-W4>14$ (red).
      }
      \label{fig:nz_reddening}
    \end{figure}
    
    \subsection{Extremely red quasar criterion}
    \label{sec:erq_criteria}
    \citet{Dey08} discovered a population of $z\sim2$ dust-obscured
    galaxies, revealed by the {\it Spitzer} Space Telescope. In addition
    to having AGN activity, these objects are strongly star forming and
    dusty which contributes to their red colours \citep{Brand08, Soifer08,
      Pope08, Bussmann09, Bussmann09b, Melbourne09, Melbourne12}. Motivated
    by this finding, we apply similar selection criteria to the
    SDSS/BOSS+WISE matched quasars; see Figure~\ref{fig:W4_vs_rband}. At
    high redshift our WISE $W4$ i.e., 22$\mu$m-based selection, is
    sensitive to hot dust since $W4$ probes $\lambda\simeq7.3\mu$m at
    $z\approx2$ \citep{Draine_Li07, Elitzur08, Diamond-Stanic10}. This is
    of particular interest because at these redshifts shorter wavebands
    (e.g. $W1$) may cease to trace dust emission, i.e. $W1$ probes
    $\lambda\simeq1\mu$m, where dust does not emit strongly.
    
    In particular, we require the spectroscopic quasars to have a reliable
    detection in the WISE $W4$-band:
    \begin{equation}
      {\rm SNR}_{W4}\geq3, 
      \label{eqn:w4_snr}
    \end{equation}
    a colour of 
    \begin{equation}
      r_{\rm AB} - W4_{\rm Vega}>14.0 
    \end{equation}
    i.e., $r_{\rm AB} - W4_{\rm AB}\gtrsim 7.5$, which corresponds to
    $F_\nu({\rm 22\mu m})/F_\nu(r) \gtrsim 1000$, and
    \begin{equation}
      W4_{\rm Vega} \leq 8.00.
      \label{eqn:w4mpro}
    \end{equation} 
    We further include objects with $W4 > 8.00$ if the object has
    SNR$_{W4}\geq3$ and $W4 - \sigma_{W4}< 8.00$. In
    Table~\ref{tab:ERQ_key_numbers} we outline the stages in the selection
    and the resulting number of quasars after every step.
    
    The $r-W4>14$ selection has been shown to select Type 2 quasars
    with $z\sim2$ \citep{Brand06, Brand07} and while both \citet{Dey08}
    and our selection have a colour cut of $r_{\rm AB} - W4_{\rm
      Vega}>14.0$, we note that \citet{Dey08} went to much fainter
    magnitudes, basing their initial selection in the IR, matched to
    optical catalogues, and then obtained additional spectra of their
    candidates. Also, while \citet{Dey08} start from the IR selection, we
    start with UV/visible selected quasars in SDSS/BOSS and then look at
    $W4$ to find the reddest cases.
    
    Our selection applied to both the AllSky and AllWISE catalogues
    results in very similar samples, and any object that passes the
    criteria in equations~(\ref{eqn:w4_snr})-(\ref{eqn:w4mpro}) in either
    of these data releases is included. There are 65 quasars that have
    spectra from SDSS/BOSS and satisfy the extremely red criterion from
    the union between the two WISE datasets.
    
    The $W4$ PSF has a full width at half maximum (FWHM) of 12$''$,
    much larger than the optical PSF, so a nearby galaxy and a quasar
    cannot be deblended in $W4$ observations if they are only a few
    arcseconds apart. To check this possibility, we visually inspected the
    SDSS and WISE images of the 65 objects using the SDSS Image List Tool
    and the IRSA WISE Image Service, and found that there is no other
    obvious optical source (down to the 5$\sigma$-depth of the SDSS
    photometric survey of $r=22.7$) within 6$''$ of any of our targets. In
    only two cases there is potential concern that another optical object
    boosts the $W4$ flux - we return to this in
    Section~\ref{sec:starburst_egs}. We do note, however, that dusty
    starforming galaxies (particularly at high redshift) are very faint
    optically, so there is still a possibility that there is an additional
    contaminating galaxy; inspecting the relatively shallow SDSS images
    only eliminates the possibility of opitcally bright contaminating
    galaxies.
    
    Also, although they are bright in $W1$ and $W2$, brown dwarfs tend
    to be significantly fainter than 8th magnitude in $W4$
    \citep{Kirkpatrick_JDavy11}, so we do not suspect contamination from
    these objects. Thus we consider all 65 $r-W4>14$ quasars to be good
    matches.

    \subsection{Effects of dust extinction}\label{sec:redd}
    Before we discuss in detail the observed spectral properties of our sample
    of extremely red quasars, we describe our expectations for how much
    extinction would be required to redden the overall spectral energy
    distribution of a quasar in order for it to be picked up by our colour
    selection. We acknowledge that the extremely red colours of our objects 
    may not be all due to dust, and thus this exercise is for general guidance. 

    We simulate the colours of reddened quasars at different redshifts
    using the \citet{Richards06b} spectral energy distribution of bright
    quasars, assuming that the spectral energy distributions (SEDs) of all
    unextincted quasars are the same. We then apply varying amounts of
    SMC-like extinction (parametrized by the amount of extinction in the
    V-band, $A_{V}$, in magnitudes), place this object at different
    redshifts and record its $r-W4$ colour.
    
    The result is shown in Figure \ref{fig:nz_reddening}, where the
    amount of extinction as a function of redshift is shown to generate
    colours of $r-W4>13$, $r-W4>13.5$ and $r-W4>14$ (the yellow, orange
    and red curves, respectively). The amount of extinction is very
    sensitive to the redshift of the quasar, and here we are only
    concerned with the dust reddening of a Type 1 SED. If there is any
    optical emission that is not affected by the circumnuclear dust
    extinction -- for example, the extended emission from the host galaxy
    -- then it would make a reddened quasar appear less red in the $r-W4$
    colour. Thus the values of $A_V$ plotted in
    Figure~\ref{fig:nz_reddening} should be regarded as {\it lower} limits
    on the amount of quasar extinction to produce the requisite
    colours. At low redshifts, the observed $r-$band probes rest-frame
    optical emission of the quasar, so the amount of extinction necessary
    to produce a very red optical-to-infrared colour is
    considerable. However, at high redshifts (where BOSS observations
    probe the rest-frame ultra-violet emission), an $A_V$ of just a few
    tenths is sufficient to strongly suppress the ultra-violet emission
    while having a small effect on the infrared; this produces a very red
    $r-W4$ colour. \citet{Vardanyan14} discuss the rest
    0.25$\mu$m/7.8$\mu$m luminosities and properties of $1.4 < z < 5$
    quasars in the SDSS, and define `obscured' quasars based on $\gtrsim$5
    mags of UV extinction; considerably larger than we consider for our
    extremely red quasars over the same redshift range.
    
    Due to the shallow nature of the WISE $W4$ band, at high redshift,
    any $W4$-detected galaxy that does not show nuclear activity will
    have $L_{\rm IR} > 10^{12} L_{\odot}$ and star-formation rates $> 100
    M_{\odot}$ yr$^{-1}$. Indeed, many of the non-AGN WISE $W4$ detections
    will have substantially higher infrared luminosities and SFRs.
    
    We have also checked the $R-W4$ colours of galaxies from the
    Canada-France-Hawaii Telescope Legacy
    Survey\footnote{\href{http://www.cfht.hawaii.edu/Science/CFHLS/}{http://www.cfht.hawaii.edu/Science/CFHLS/}.}
    and find that they again have a range of colours, $\sim$~$9<R-W4$~$\lesssim15$.
    
    The optical colours of our sample are inconsistent with simple
    dust extinction of a standard quasar SED \citep[e.g.][]{Richards03,
      Krawczyk15} and this is a current area of investigation. We discuss
    possible solutions to this in our companion papers, Hamann et
    al. (2015, in prep.) and Zakamska et al. (2015, in prep); in
    particular, contribution from scattered light is a possibility
    \citep[e.g., ][]{Zakamska05, Zakamska06} and so is patchy obscuration
    \cite{Veilleux13}.  In particular, we investigate further the nature
    of the obscuration from the SEDs, which is not consistent with simple
    reddening from our analysis. As such, we are very hesitant to draw any
    futher conclusions on dust reddening and $A_{V}$ values in this paper,
    and leave detailed discussions to Hamann et al. (2015, in prep).

{\onecolumn
  \begin{landscape}
    \scriptsize
   \begin{longtable}{l rrr rr rr rr ll}  
      \captionsetup{width=18cm}
      \caption{The 65 quasars from SDSS and BOSS that have $r-W4>14$, in
        order of increasing redshift.  The {\it r}-band AB magnitude is from
        either \citet{Schneider10} or \citet{Paris14}, corrected for Galactic 
        extinction \citep{Schlegel98}, while the WISE W4 Vega magnitude is
        from the
        \href{http://wise2.ipac.caltech.edu/docs/release/allwise/}{AllWISE
          Data Release catalogue}. The Radio fluxes are the integrated flux
        densities ($F_{\rm int}$) measured in mJy at 20cm from the FIRST
        survey \citep{Becker95}. The Rest-frame Equivalent Widths (REWs) 
        in \AA ngstroms are given for \mgii at
        $z\leq1.5$ and \civ for objects at redshift $z>1.5$. The Full-Width,
        Half-Maximum (FWHM) of the relevant emission line is also quoted. 
        Objects with similar optical spectral properties are grouped together, 
        using the bitwise values described in Table~\ref{tab:ERQ_bitmask}, converted
        here to binary (with a space gap for easier inspection).}\\
      \hline 
      \hline 
      Name      & \multicolumn{1}{c}{R.A.}      & \multicolumn{1}{c}{Decl.}    &  \multicolumn{1}{c}{$z$} & \multicolumn{1}{c}{{\it r}-mag} & \multicolumn{1}{c}{$W4$ mag} &  $r-W4$  &  Radio & REWs   & FWHM                &  \multicolumn{1}{c}{Notes}  & \multicolumn{1}{c}{Spectral} \\ 
      (SDSS~J) &  \multicolumn{1}{c}{(J2000)} & \multicolumn{1}{c}{(J2000)} &                                &  \multicolumn{1}{c}{(AB)}        & \multicolumn{1}{c}{(Vega)}     &                &    flux/mJy   &  \multicolumn{1}{c}{ / \AA\ } & / km s$^{-1}$ &  & \multicolumn{1}{c}{Classification}      \\
      \hline
      \endfirsthead
      \hline 
      \hline 
      Name      & \multicolumn{1}{c}{R.A.}      & \multicolumn{1}{c}{Decl.}    &  \multicolumn{1}{c}{$z$} & \multicolumn{1}{c}{{\it r}-mag} & \multicolumn{1}{c}{$W4$ mag} &  $r-W4$  & Radio & REWs   & FWHM                &  \multicolumn{1}{c}{Notes}  & \multicolumn{1}{c}{Spectral} \\ 
      (SDSS~J) &  \multicolumn{1}{c}{(J2000)} & \multicolumn{1}{c}{(J2000)} &                                &  \multicolumn{1}{c}{(AB)}        & \multicolumn{1}{c}{(Vega)}     &                &     flux/mJy  &  \multicolumn{1}{c}{ / \AA\ } & / km s$^{-1}$ &   &  \multicolumn{1}{c}{Classification}    \\
      \hline
      \hline
      \endhead
      \hline
      \endfoot
       101034.27+372514.7   &  152.64283 &    37.42077 &   0.282 &   18.39$\pm$0.03   &   4.26$\pm$0.02   &  14.14 &     -  &          - &             - & Type 1; Strong narrow [OII], starburst                     & 0000 0101 \\
 0941000.8+143614.4   &  145.25339 &    14.60402 &   0.384 &   19.43$\pm$0.02   &   5.39$\pm$0.04   &  14.04 &    7.6 &          - &             - & Type 1; Strong narrow [OII], starburst     	       	 & 0001 0110 \\
 105327.92+375804.2   &  163.36636 &    37.96786 &   0.448 &   19.64$\pm$0.03   &   5.59$\pm$0.04   &  14.05 &    2.8 &          - &             - & Type 1; Weak [OII], strong [NeIII] and [OIII], post-starburst     & 0001 1010 \\
 112657.76+163912.0   &  171.74069 &    16.65334 &   0.464 &   19.20$\pm$0.02   &   4.79$\pm$0.03   &  14.41 &    2.4 &          - &             - & Type 1; Strong narrow [OIII], post-starburst      	         & 0001 1010 \\
 102834.03-023659.6   &  157.14182 &    -2.61657 &   0.470 &   18.90$\pm$0.02   &   4.80$\pm$0.03   &  14.10 &   68.5 &          - &             - & Type 1 based on MgII, blazar candidate	 & 0001 0001 \\
 110550.54+112702.0   &  166.46059 &    11.45058 &   0.498 &   20.65$\pm$0.05   &   6.17$\pm$0.06   &  14.48 &    2.9 &          - &             - & Potential Type 2 candidate, but broad H$\alpha$    & 0001 0010 \\
\hline
 092501.78+274607.9   &  141.25744 &    27.76888 &   0.531 &   19.43$\pm$0.02   &   4.47$\pm$0.02   &  14.97 &   26.7 &          - &             - & Type 2 in \citet{Reyes08}                         	 & 0001 0010 \\
 091157.55+014327.5   &  137.98980 &     1.72432 &   0.603 &   21.48$\pm$0.08   &   6.60$\pm$0.09   &  14.88 &    4.6 &          - &             - & Type 2 in \citet{Reyes08}                           & 0001 0010 \\
 225612.16-010508.0   &  344.05071 &    -1.08556 &   0.651 &   21.53$\pm$0.07   &   7.16$\pm$0.13   &  14.37 &    3.0 &  27$\pm$8  & 1870$\pm$730  & Type 2 in \citet{Reyes08}                           & 0001 0010 \\
 161558.92+252736.1   &  243.99552 &    25.46004 &   0.676 &   22.04$\pm$0.09   &   7.26$\pm$0.09   &  14.78 &    4.2 &  69$\pm$8  &  918$\pm$174  & Type 2    	                    		 & 0001 0010 \\
 111354.66+124439.0   &  168.47776 &    12.74418 &   0.680 &   19.86$\pm$0.03   &   5.20$\pm$0.04   &  14.66 &    4.2 &          - &             - & Red Type 1 (broad H$\beta$, strong [OII], Fe II)	 & 0001 0101 \\
 124836.10+424259.3   &  192.15043 &    42.71650 &   0.682 &   21.61$\pm$0.07   &   7.32$\pm$0.10   &  14.29 &    3.4 &  22$\pm$11 & 3103$\pm$1138 & Type 2  	                                         & 0001 0010 \\
\hline
 101303.09+072731.9   &  153.26288 &     7.45887 &   0.697 &   20.99$\pm$0.05   &   6.91$\pm$0.10   &  14.08 &     -  &  35$\pm$5  & 1793$\pm$294  & Type 2  	                                         & 0000 0010 \\
 154215.25+161648.9   &  235.56354 &    16.28028 &   0.714 &   21.70$\pm$0.06   &   7.54$\pm$0.10   &  14.17 &    1.5 &  23$\pm$6  & 1972$\pm$507  & Type 2  	                                         & 0001 0010 \\
 144642.28+011303.0   &  221.67618 &     1.21751 &   0.726 &   21.02$\pm$0.05   &   6.08$\pm$0.05   &  14.94 &    5.8 &        -   &            -  & Type 2 in \citet{Reyes08}                          & 0001 0010 \\
 005009.81-003900.6   &   12.54091 &    -0.65017 &   0.728 &   20.11$\pm$0.03   &   5.87$\pm$0.06   &  14.24 &    4.4 &        -   &            -  & Type 2    			                 & 0001 0010 \\
 142240.99+383710.4   &  215.67083 &    38.61957 &   0.758 &   20.57$\pm$0.03   &   6.53$\pm$0.05   &  14.04 &    9.8 &        -   &            -  & Type 2 in \citet{Reyes08}                    & 0001 0010 \\
 140744.00+360109.5   &  211.93333 &    36.01931 &   0.783 &   21.29$\pm$0.05   &   6.85$\pm$0.07   &  14.44 &    5.3 & 112$\pm$5  &    (4000)     & Type 1 with O[II]		                         & 0001 0101 \\
\hline
 124231.04+023307.0   &  190.62933 &     2.55195 &   0.785 &   21.16$\pm$0.05   &   6.81$\pm$0.08   &  14.35 &     -  &         -  &             - & Red Type 1  	                                 & 0000 0001 \\
 091501.71+241812.1   &  138.75714 &    24.30336 &   0.844 &   20.39$\pm$0.03   &   4.82$\pm$0.03   &  15.57 &   10.1 &  -         &             - & Type 1; Very bright in $W4$                             & 0001 0001 \\
 152504.74+123401.7   &  231.26977 &    12.56715 &   0.851 &   21.14$\pm$0.04   &   6.92$\pm$0.07   &  14.22 &     -  &  43$\pm$3  &  1984$\pm$167 & Type 2	    	                                 & 0000 0010 \\
 151354.48+145125.2   &  228.47702 &    14.85702 &   0.881 &   21.00$\pm$0.04   &   6.37$\pm$0.05   &  14.63 &    7.6 &  20$\pm$2  &  2212$\pm$209 & Type 2  	                                         & 0001 0010 \\
 144855.23+213622.0   &  222.23016 &    21.60612 &   0.882 &   21.70$\pm$0.07   &   7.72$\pm$0.12   &  13.98 &    5.2 & 101$\pm$17 & 7868$\pm$1347 & Type 2 	                                         & 0001 0010 \\
 210712.77+005439.4   &  316.80322 &     0.91096 &   0.924 &   20.56$\pm$0.03   &   6.31$\pm$0.06   &  14.25 &    0.5 &          - &             - & Type 1, strong FeII, LoBAL MgII absorption         & 0011 0001 \\
\hline
 075257.36+351834.6   &  118.23901 &    35.30962 &   0.926 &   23.23$\pm$0.30   &   7.57$\pm$0.19   &  15.66 &     -  &  52$\pm$4  &  2979$\pm$198 & Type 1         	    	 	                 & 0000 0001  \\
 141702.32+364839.5   &  214.25969 &    36.81098 &   1.004 &   21.58$\pm$0.06   &   6.89$\pm$0.06   &  14.69 &    7.8 & 181$\pm$33 & 21698$\pm$3680 & Type 2  				                 & 0001 0010  \\
 142131.51+210719.4   &  215.38131 &    21.12207 &   1.061 &   21.78$\pm$0.08   &   7.72$\pm$0.14   &  14.06 &    1.0 &  37$\pm$4  &  1425$\pm$137 & Type 2 	                                         & 0001 0010  \\
 092356.70+230030.9   &  140.98627 &    23.00860 &   1.095 &   22.34$\pm$0.12   &   7.68$\pm$0.20   &  14.66 &     -  &  30$\pm$8  &  2112$\pm$410 & Type 2, strong O[II]	                         & 0000 0110  \\ 
 091103.50+444630.3   &  137.76458 &    44.77511 &   1.302 &   20.88$\pm$0.06   &   6.58$\pm$0.06   &  14.30 &     -  &          - &             - & Type 1; Strong FeII emission, LoBAL MgII absorption        & 0010 0001  \\
 111004.78+375236.7   &  167.51994 &    37.87686 &   1.308 &   20.34$\pm$0.04   &   5.61$\pm$0.04   &  14.74 &    7.1 &          - &             - & Red Type 1 	       		                 & 0001 0001  \\
\hline
 165258.51+390249.7   &  253.24380 &    39.04716 &   1.308 &   21.21$\pm$0.04   &   6.79$\pm$0.06   &  14.42 &  249.3 &   7$\pm$1   &  2966$\pm$304  & Type 1 based on MgII, blazar candidate       & 0001 0001 \\
 102425.41+033239.8   &  156.10589 &     3.54441 &   1.477 &   22.40$\pm$0.17   &   7.11$\pm$0.17   &  15.29 &     -  &   35$\pm$8  &  1166$\pm$377  & Type 2 based on v. narrow CIV 	                 & 0000 0010 \\
 144929.61+394824.2   &  222.37341 &    39.80674 &   1.491 &   21.74$\pm$0.06   &   7.66$\pm$0.11   &  14.08 &   24.3 &   57$\pm$9  & 10075$\pm$1405 & Type 2 based on v. narrow CIV and HeII		 & 0001 0010 \\
 153542.41+090341.1   &  233.92671 &     9.06143 &   1.533 &   21.33$\pm$0.04   &   6.65$\pm$0.07   &  14.68 &    8.1 &  137$\pm$14 &  1556$\pm$140  & Type 1; Very strong FeII emission in UV2/3, UV1,      & 0011 0001 \\
                      &            &             &         &                    &                   &        &        &             &                & and strong AlIII with weak/absent CIII] and SiIII]; see Appendix~\ref{sec:J1535}.    &           \\ 
 080304.76+532627.0   &  120.76987 &    53.44084 &   1.573 &   20.85$\pm$0.05   &   6.60$\pm$0.07   &  14.26 &    2.9 &  123$\pm$7  &  3196$\pm$138  & Type 1; LoBAL	      	       	  	      	 & 0011 0001 \\	
 122453.20+084303.2   &  186.22167 &     8.71757 &   1.960 &   23.55$\pm$0.73   &   8.01$\pm$0.25   &  15.54 &    -   &   0$\pm$0   &      0$\pm$0   & Red Type 1 	                                 & 0000 0001 \\	
\hline
 211329.61+001841.7   &  318.37340 &     0.31161 &   1.996 &   23.40$\pm$0.28   &   8.14$\pm$0.29   &  15.26 &    -  & 192$\pm$7   &  1536$\pm$49   & Type 2, ``Extreme'' REW (EREW)                    & 1000 0010   \\ 
 163313.29+401338.9   &  248.30541 &    40.22749 &   2.010 &   21.18$\pm$0.05   &   7.12$\pm$0.08   &  14.07 &  10.5 &  20$\pm$10  & 11936$\pm$99 & Type 1; Blazar candidate                                  & 0001 0001   \\
 173049.10+585059.5   &  262.70459 &    58.84988 &   2.035 &   21.39$\pm$0.07   &   7.37$\pm$0.09   &  14.02 &    -  &          -  &              - & Type 1; FeLoBAL    		    	                 & 0010 0001   \\	
 133611.79+404522.9   &  204.04913 &    40.75637 &   2.071 &   21.07$\pm$0.06   &   6.38$\pm$0.05   &  14.69 &   3.1 &  20$\pm$3   &  6244$\pm$688  & Type 1				                 & 0001 0001   \\	
 013435.66-093102.9   &   23.64861 &    -9.51748 &   2.220 &   21.30$\pm$0.06   &   6.58$\pm$0.07   &  14.71 &   960 &         -   &              - & Gravitational Lens, Red Type 1, see Appendix~\ref{sec:J0134}      & 0001 0001   \\ 
 084447.66+462338.7   &  131.19862 &    46.39411 &   2.224 &   21.48$\pm$0.08   &   7.48$\pm$0.13   &  14.00 &    -  &   175$\pm$4 &  1744$\pm$34  & Type 2, with narrow CIV 	                         & 0000 0010   \\	
\hline
 094317.59+541705.1   &  145.82333 &    54.28475 &   2.230 &   21.41$\pm$0.07   &   7.33$\pm$0.09   &  14.08 &   2.2 &           - &             - & Type 1; FeLoBAL 	  	 	                         & 0011 0001   \\ 
 222646.53+005211.2   &  336.69390 &     0.86980 &   2.247 &   21.54$\pm$0.07   &   7.54$\pm$0.15   &  14.00 &  653  &   65$\pm$4  &  2663$\pm$212 & Red Type 1, strong associated absorbtion               & 0001 0001   \\ 
 000610.67+121501.2   &    1.54449 &    12.25035 &   2.309 &   22.18$\pm$0.13   &   6.56$\pm$0.07   &  15.61 &    -  &  108$\pm$6  &  4589$\pm$202 & Type 1, EREW	    	      	                 & 1000 0001   \\	
 232326.17-010033.1   &  350.85906 &    -1.00920 &   2.355 &   21.83$\pm$0.08   &   7.76$\pm$0.22   &  14.07 &    -  &  256$\pm$5  &  3977$\pm$62  & Type 1, N\,{\sc v}/\lya$>1$, EREW                  & 1000 0001   \\	
 123241.73+091209.3   &  188.17391 &     9.20261 &   2.385 &   21.09$\pm$0.05   &   6.78$\pm$0.09   &  14.31 &    -  &  230$\pm$3  &  4873$\pm$53  & Type 1, N\,{\sc v}/\lya$>1$, EREW                  & 1000 0001   \\	
 083413.90+511214.6   &  128.55793 &    51.20407 &   2.391 &   19.77$\pm$0.04   &   5.43$\pm$0.04   &  14.33 &   1.9 &        -    &            -  & Type 1, red slope, $z$ from absorption at red end  & 0001 0001  \\
\hline
 0832000.2+1615000.3  &  128.00085 &    16.25011 &   2.423 &   21.82$\pm$0.08   &   7.51$\pm$0.16   &  14.31 &   1.1 &        -    &            -  & Type 1, EREW, strong associated absorption lines   & 1001 0001  \\
 011110.83+204543.8   &   17.79517 &    20.76218 &   2.432 &   22.03$\pm$0.09   &   7.96$\pm$0.21   &  14.06 &    -  &  0$\pm$0    &    0$\pm$0    & Type 1; Ly-$\alpha$ emitter QSO 	  	    	         & 0000 0001  \\ 
 090630.73+082837.3   &  136.62806 &     8.47705 &   2.437 &   21.77$\pm$0.09   &   7.76$\pm$0.20   &  14.00 &  22.8 &  3$\pm$1    &  340$\pm$159  & Type 1 (from Ly-$\alpha$ width)                    & 0001 0001  \\	
 093638.41+101930.3   &  144.16006 &    10.32510 &   2.458 &   21.72$\pm$0.06   &   7.61$\pm$0.17   &  14.11 &    -  & 169$\pm$4   & 1273$\pm$20   & Type 2, but with EREW	                         & 1000 0010  \\	
 134026.99+083427.2   &  205.11248 &     8.57424 &   2.490 &   22.31$\pm$0.13   &   8.10$\pm$0.21   &  14.21 &    -  & 338$\pm$11  & 2088$\pm$51   & Type 2, but with EREW                              & 1000 0010  \\ 
 221524.00-005643.8   &  333.85003 &    -0.94550 &   2.493 &   22.29$\pm$0.12   &   7.91$\pm$0.24   &  14.38 &    -  &  152$\pm$5  &  4226$\pm$112  & Type 1; EREW	                                         & 1000 0001  \\	
\hline
 111017.13+193012.5   &  167.57139 &    19.50347 &   2.497 &   20.61$\pm$0.03   &   6.59$\pm$0.07   &  14.01 &    -  &         -   &             -  & Type 1; EREW    	                                         & 1000 0010  \\ 
 155102.79+084401.1   &  237.76163 &     8.73366 &   2.520 &   20.66$\pm$0.03   &   6.61$\pm$0.06   &  14.05 &   2.9 &   10$\pm$3  &  2467$\pm$426  & Type 1; FeLoBAL	                                         & 0011 0001  \\	
 083448.48+015921.1   &  128.70202 &     1.98921 &   2.594 &   21.19$\pm$0.05   &   6.88$\pm$0.09   &  14.31 &    -  &  214$\pm$7  &  2864$\pm$73   & EREW, Type 1                                       & 1000 0001  \\ 
 220337.79+121955.3   &  330.90749 &    12.33204 &   2.626 &   21.57$\pm$0.05   &   7.45$\pm$0.14   &  14.12 &    -  &  262$\pm$3  &  1072$\pm$10   & EREW, Type 2 	                                 & 1000 0010  \\	 
 085124.78+314855.7   &  132.85329 &    31.81548 &   2.638 &   21.63$\pm$0.07   &   6.79$\pm$0.08   &  14.84 &    -  &   76$\pm$5  &  3584$\pm$181  & Type 1, W1W2-dropout                                  & 0100 0001  \\	
 100424.88+122922.2   &  151.10370 &    12.48952 &   2.640 &   22.15$\pm$0.12   &   7.87$\pm$0.19   &  14.28 &  12.3 &          -  &            -   & Tpye FeLoBAL	                                         & 0011 0001  \\	
\hline
 104611.50+024351.6   &  161.54796 &     2.73103 &   2.772 &   21.51$\pm$0.07   &   7.11$\pm$0.11   &  14.40 &    -  &   203$\pm$7 & 5181$\pm$144   & Type 1, N\,{\sc v}/\lya$>1$, EREW                  & 1000 0001  \\ 
 155434.17+110950.6   &  238.64238 &    11.16407 &   2.936 &   21.19$\pm$0.05   &   6.63$\pm$0.08   &  14.56 &    -  &          -  &            -   & Type 1, BAL, probably FeLoBAL	                         & 0010 0001  \\ 
 135959.73+052512.3   &  209.99888 &     5.42008 &   3.055 &   22.08$\pm$0.11   &   7.27$\pm$0.09   &  14.81 &    -  &   70$\pm$10 &  7250$\pm$833  & Type 1, W1W2-dropout                                       & 0100 0001  \\	
 022052.11+013711.1   &   35.21715 &     1.61976 &   3.138 &   21.82$\pm$0.08   &   7.08$\pm$0.09   &  14.74 &    -  &  341$\pm$19 &  2843$\pm$167  & EREW, Type 1, W1W2-dropout	                         & 0100 0001  \\	 
 101439.51+413830.6   &  153.66466 &    41.64183 &   4.360 &   21.84$\pm$0.08   &   7.74$\pm$0.16   &  14.10 &    -  &  118$\pm$8  & 13568$\pm$684  & Type 1, (Fe)LoBAL		                                 & 0010 0001  \\	
      \hline  
      \hline 
    \end{longtable}
    \normalsize
  \end{landscape}
}
\twocolumn

\begin{figure}
  \includegraphics[width=8.50cm, height=8.00cm, trim=0.2cm 0.2cm 0.0cm 0.0cm, clip]
  {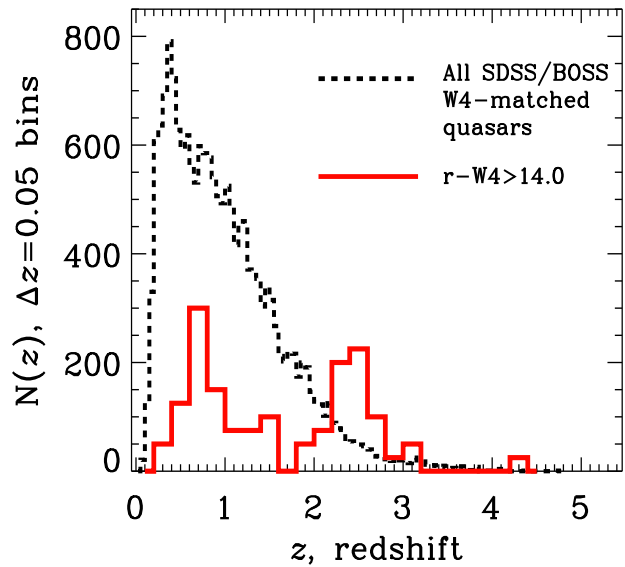}
  \centering
  \vspace{-16pt}
  \caption[]{The redshift distribution of the 17,744 SDSS/BOSS quasars with
    good W4 matches is given by the black (dashed) histogram. The redshift
    distribution of extremely red quasars (with $r-W4>14$) is given by the
    solid (red) line. It has a binwidth of $\Delta z=0.20$ and is renormalized to peak
    at 300. This maximum bin, at $z\approx0.70$, actually has 12 objects.}
 \label{fig:ERQ_Nofz}
\end{figure}
\section{Key Properties of the Extremely Red Quasars}
\label{sec:key_props}

    \subsection{Magnitudes and redshift distribution}
    The key properties of the extremely red quasar sample are provided
    in Table~2: the catalogue name of the object, its Right Ascension,
    Declination, redshift, and magnitude and colour properties.  The Radio
    fluxes are the integrated flux densities ($F_{\rm int}$) measured in
    mJy at 20cm from the FIRST survey \citep{Becker95}. The Rest-frame
    Equivalent Widths (REWs) are given for \mgii at $z\leq1.5$ and \civ
    for objects at redshift $z>1.5$. The Full-Width, Half-Maximum (FWHM)
    of the relevant emission line is also quoted, with the REW and FWHM
    measurements coming from fits to the line profiles, as described in
    Hamann et al. (in prep.).
    
    \begin{table}
      \begin{center}
        \begin{tabular}{lrrr}
          \hline
          \hline
          General         &  Bitwise & Number in \\
          Description  &   Value   & Sample \\
          \hline
          Type 1 Quasar       & 0000 0001   &  37  \\   
          Type 2 Quasar       & 0000 0010   &  28  \\    
          \hline
          Starburst	           & 0000 0100   &    5   \\  
          post-starburst      & 0000 1000   &    2   \\	
          Radio Detected	   & 0001 0000   &  37  \\    
          (Fe)LoBAL              & 0010 0000   &  10  \\   
          W1W2-dropouts   & 0100 0000   &    3  \\   
          ``Extreme'' REWs & 1000 0000   &  12  \\   
          \hline
          \hline
        \end{tabular}
        \caption{The classifications of the optical spectra of the extremely red
          quasar sample, the bitwise value associated with each classification
          and the number in the sample. Aside from the Type 1/Type 2 dichotomy,
          the categories are not mutually exclusive. Quasars are flagged as `Radio Detected' if 
          they are detected in the FIRST survey.}
        \label{tab:ERQ_bitmask}
      \end{center}
    \end{table}
 
    \begin{figure*}
      \includegraphics[width=8.50cm, height=8.00cm, trim=0.6cm 0.2cm 0cm 0.0cm]
      {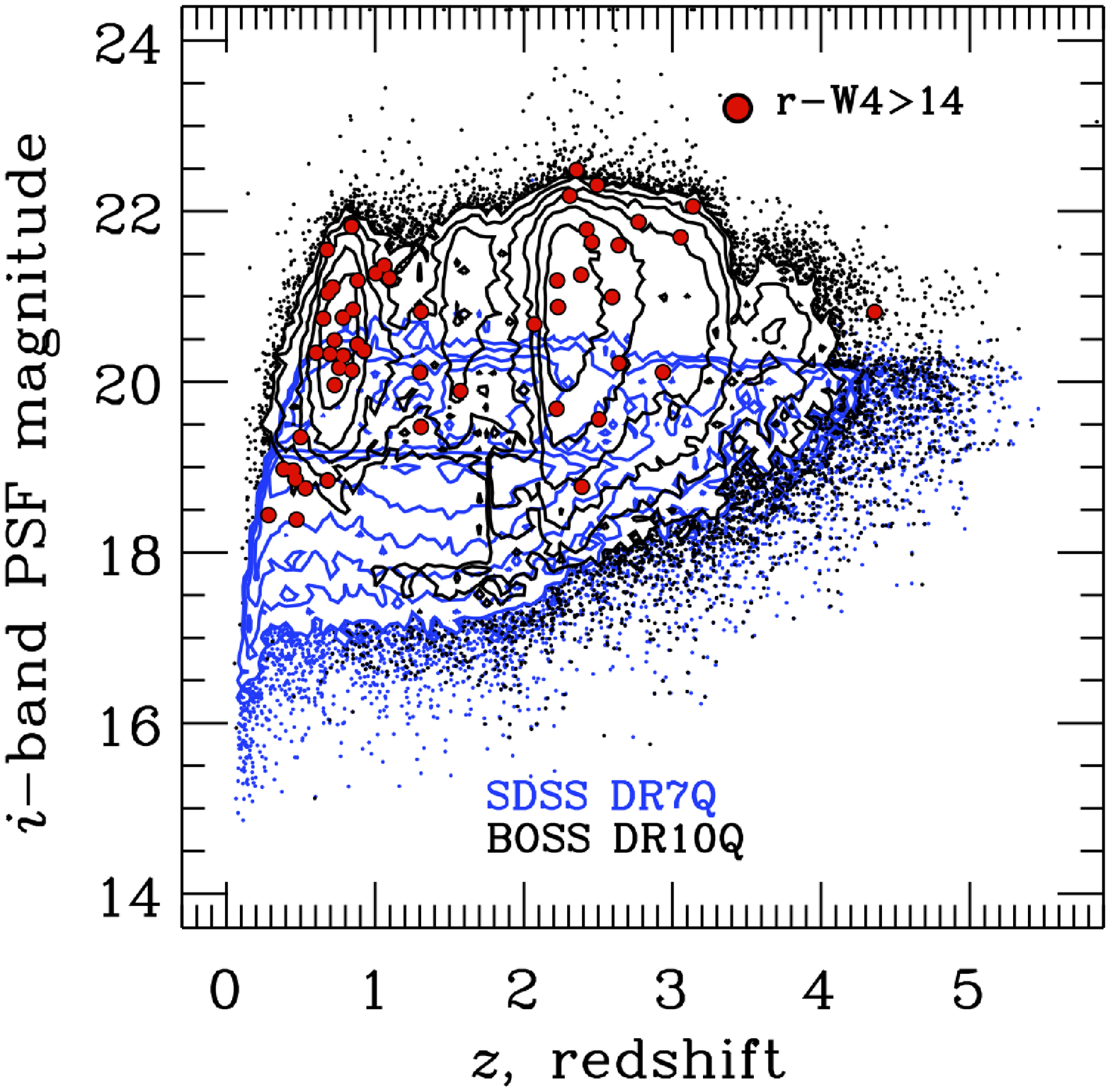}
      \includegraphics[width=8.50cm, height=8.00cm, trim=0.6cm 0.2cm 0cm 0.0cm]
      {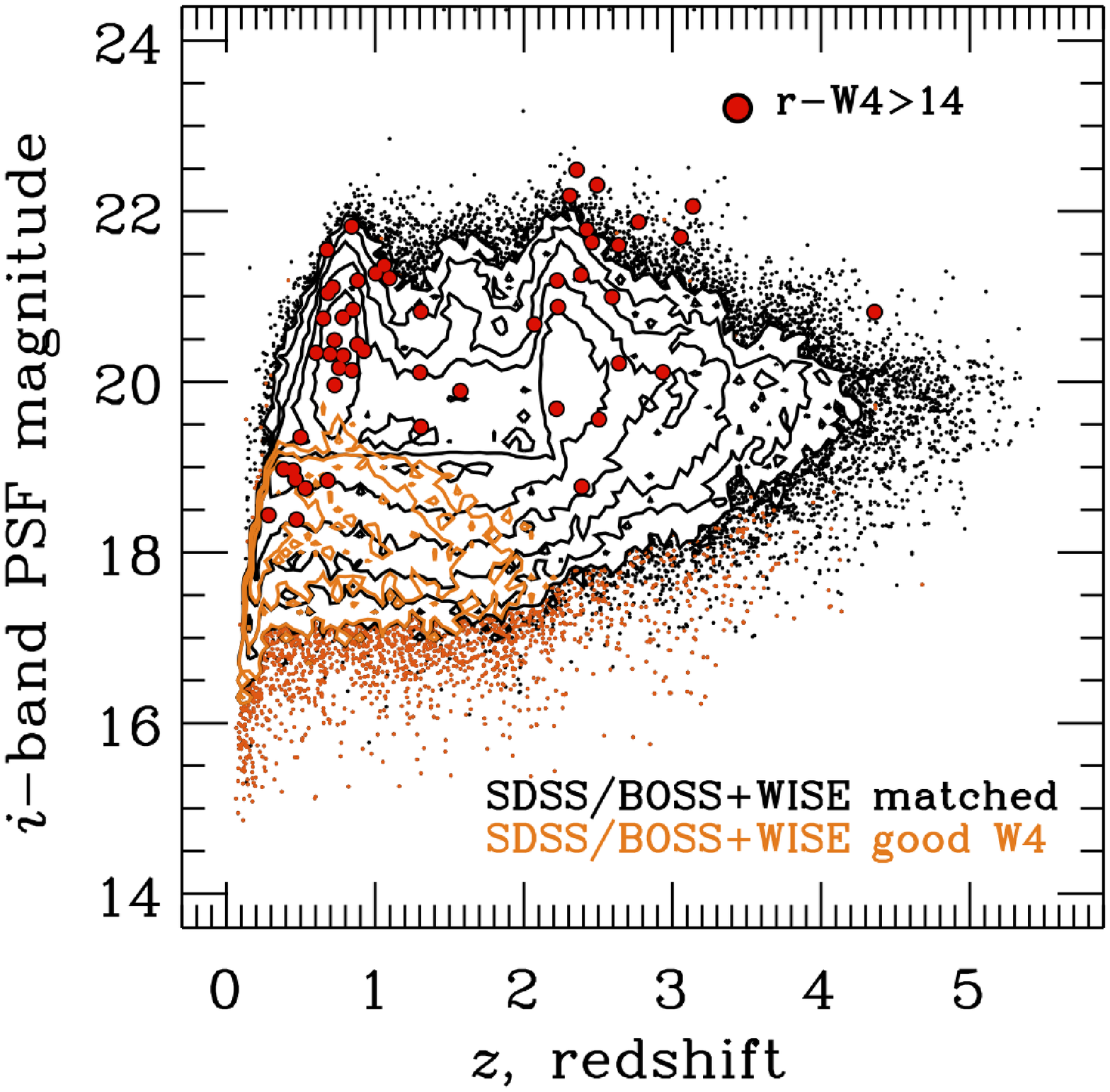}
      \centering
      \vspace{-14pt}
      \caption[]{The redshift versus $i-$band magnitude distribution for the
        extremely red quasars satisfying the $r-W4>14$ selection. {\it Left:}
        The blue and black contours show the distributions for the SDSS DR7
        and the BOSS DR10 quasar catalogues, respectively. {\it Right:} The
        black contours show the distribution of all SDSS/BOSS quasars that
        have a match in any of the WISE bands, while the orange contours are
        SDSS/BOSS quasars with good 22$\mu$m W4-band detections.}
      \label{fig:redshift_vs_iband}
    \end{figure*}

    We give each object a classification based on optical spectral
    properties, indicated by the bitwise values described in
    Table~\ref{tab:ERQ_bitmask}. These classifications were all performed
    by visual inspection, with objects assigned to the classes given in
    Table~\ref{tab:ERQ_bitmask} depending on the presence (or absence) of
    broad/narrow emission lines, broad absorption features, indications of
    ongoing or recently ceased star-formation, or objects with large
    equivalent width in the \civ emission line. These visual inspections
    are generally qualitative but as discussed in Section~4, aid us to
    understand the sample and physical types of quasars that the $r-W4>14$
    colour selects. We also present a short note on each object in Table
    2, qualitative assessments based on visual inspection of the optical
    spectra, and use these notes to make the link between the
    classifications given in Table~\ref{tab:ERQ_bitmask} and the
    discussions presented in Section~\ref{sec:spectral_egs}. We call a
    quasar ``red'' when its spectral slope is shallower than the
    \citet{VdB01} template. The acronym EREW stands for Extreme Rest-frame
    Equivalent Width, referring to a group of objects we introduce and
    describe in Section~\ref{sec:erews}.
    
    The redshift range for the extremely red quasars is
    $0.282<z<4.36$, and half the sample is fainter than
    $r$=21.4. Figure~\ref{fig:ERQ_Nofz} shows the redshift distribution of
    the sample compared to the full redshift distribution of the 17,744
    SDSS/BOSS quasars with good W4 matches. The redshift distribution of
    the $r-W4>14$ sample is bimodal, with peaks around $z\sim0.8$ and
    $z\sim2.5$. This bimodality is the result of the two different
      SDSS quasar selections. The $z\sim2.5$ redshift peak is due to the
      increased depth of the BOSS selection, as seen in
      Figure~\ref{fig:redshift_vs_iband}.
    
    In the left panel of Figure~\ref{fig:redshift_vs_iband} the
    redshifts and $i$-band magnitudes are displayed for the $r-W4>14$
    objects, and compared to the distributions from the full SDSS DR7 and
    BOSS DR10 quasar catalogues. SDSS quasars use the $i$-band for
    selection limits, thus one can identify the $i=19.1$ magnitude limit
    contour for the SDSS $z\lesssim2.5$ quasars, as well as the redshift
    histogram peaks at $z\sim0.6-1.0$ and $z\sim2.2-2.7$ seen in the BOSS
    quasar distribution \citep[e.g. Fig. 2 in][]{Paris14}. In the right
    panel of Fig.~\ref{fig:redshift_vs_iband}, the extremely red quasars
    are compared to the SDSS/BOSS+WISE catalogue, for the fully matched
    (i.e. any WISE band) sample given by the black contours, and for the
    $W4$-matched objects only, given by the orange contours. The
    redshift-$i$ magnitude distribution of the $\approx$18,000 objects
    with good $W4$ matches is given by the orange contours and
    points. These $W4$ objects are at the bright end in the optical, and
    are generally at $z<2$.

    \begin{figure}
      \includegraphics[width=8.40cm, height=8.00cm, trim=0.8cm 0.2cm 0.8cm 0.0cm]
      {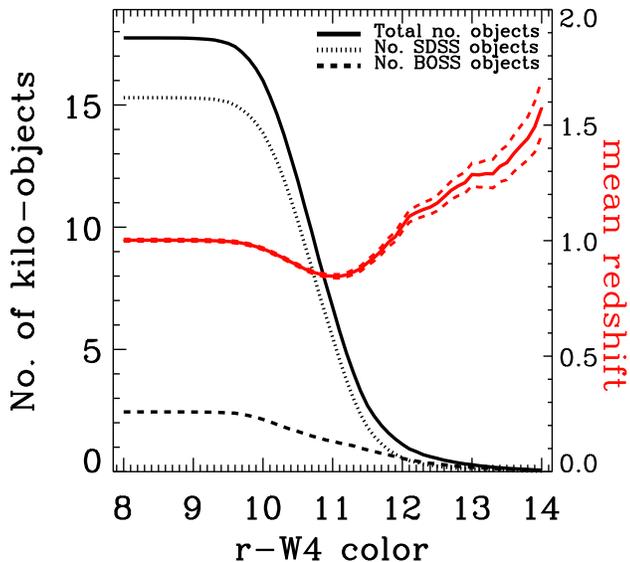}
      \centering
      \vspace{-14pt}
      \caption[]{The number of quasars selected, and their mean redshift, as a
        function of $r_{\rm AB}-W4_{\rm Vega}$ colour. The dashed (red) lines give the 
        1$\sigma$-spread on the mean redshift value. In total, 17,744
        quasars have $r-W4>8.0$.  Note that the extremely red quasar sample at
        $r-W4\geq14.0$ is the tail end of the $W4$-matched quasar
        sample.}
      \label{fig:N_vs_color}
    \end{figure}
    Since we are selecting such extreme objects it is important to see
    how the selection affects the redshift distribution. In
    Figure~\ref{fig:N_vs_color} we show how the number of objects
    selected, and the sample redshift, depend on $r-W4$ colour. The dip in
    mean redshift at $r-W4\sim10-11$ comes from the fact that fainter BOSS
    quasars, generally at $z\gtrsim2$, are lost quicker than $z\lesssim1$
    quasars. The upturn at $r-W4>11.5$ is due to the fact that the
    $W4$-bright BOSS quasars that are very red remain in the sample even
    to the reddest colours.

\begin{figure}
      \centering
      \includegraphics[width=8.60cm, height=8.60cm, trim=1.6cm 0.9cm 0.4cm 3.2cm]
      {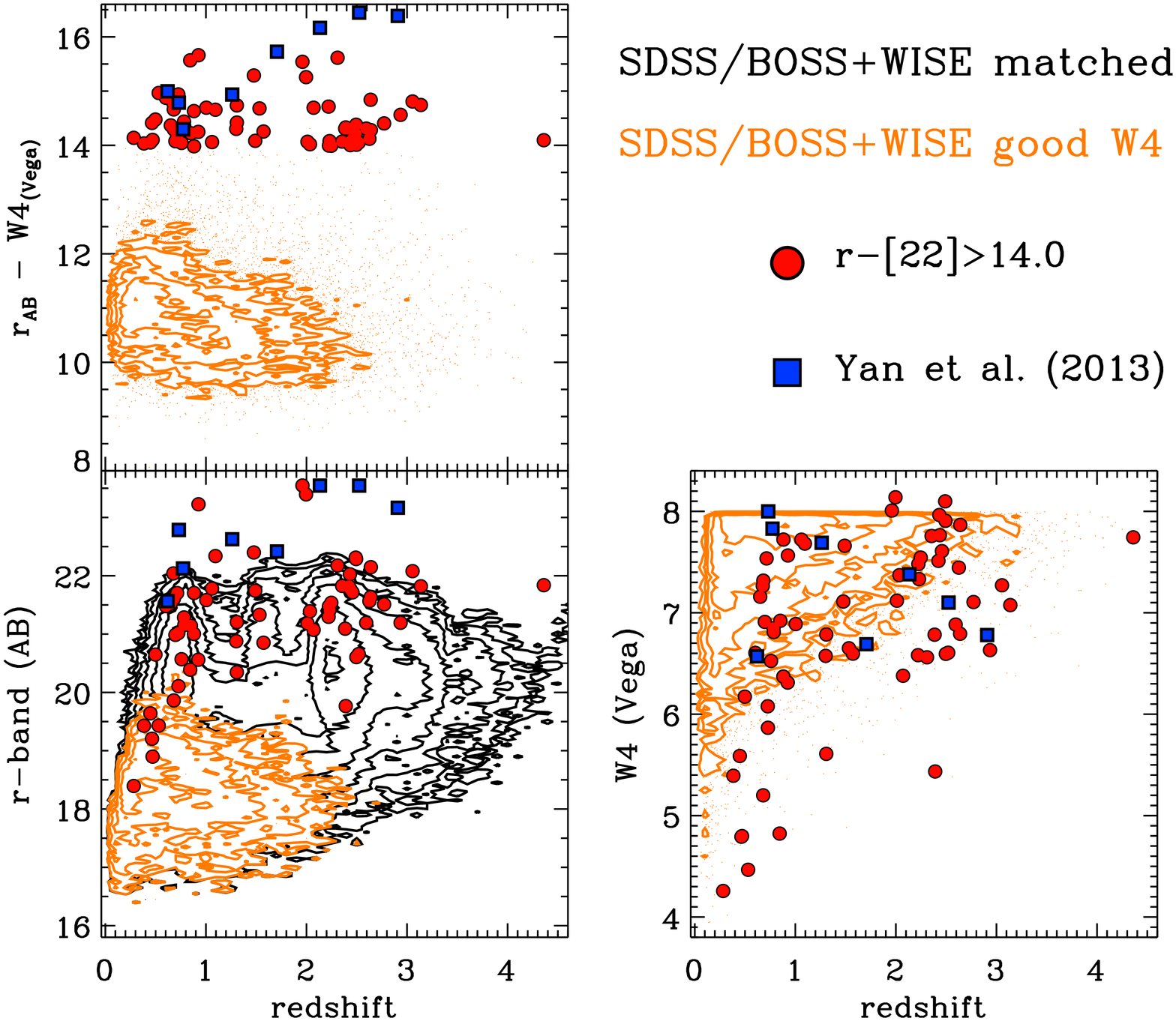}
      \vspace{-20pt}
      \caption{
        The magnitude-redshift and colour-redshift diagrams for the objects
        with a WISE detection (black contours) and objects with a good $W4$ match
        (orange contours). The $r-W4>14$ extremely red quasar population is 
        shown by the red circles.  {\it Top left:} Redshift vs. $r-W4$ colour;
        {\it bottom left:} Redshift vs. $r$-band AB magnitude and {\it bottom
          right:} Redshift vs. W4 Vega magnitude. The data from \citet{Yan13}
        are indicated by the blue squares.}
      \label{fig:ERQs_magredshift}
    \end{figure}

    Figure~\ref{fig:ERQs_magredshift} presents the magnitude-redshift
    and colour-redshift relations for the SDSS/BOSS+WISE matched catalogue
    highlighting the extremely red sample. This plot can be directly
    compared with \citet{Yan13} (and their Figure~20). \citet{Yan13} have
    also merged the WISE photometry with the SDSS (though not BOSS), and
    describe the spectroscopic observations of eight objects that have
    $r-W4>14$ (see their Table 1 for details). Although we have an order
    of magnitude more $r-W4>14$ objects at $z>2$, the Yan et al. objects
    are significantly fainter in the optical than our SDSS/BOSS detected
    objects. Based on the spectral properties of the \citet{Yan13} sample,
    3 (out of 8) objects are classified as Type-1 AGN, 3 objects are
    Type-2 AGN, and the remaining two objects have features consistent
    with either a Type-2 AGN, or a star-forming galaxy, broadly in line
    with the composition given in
    Table~\ref{tab:ERQ_bitmask}. Interestingly, \citet{Yan13} motivate the
    use of the $r-W4>14$ colour selection in order to select ultraluminous
    infrared galaxy (ULIRG) candidates, not red quasars (they use the
    shorter WISE bands to select reddened quasars), and indeed find this
    colour selection identifies IR luminous galaxies over a wide range of
    redshift, $z\sim0.7-3$. However, the relative contribution to the $W4$
    luminosity from AGN and star formation warrants investigation,
    especially at high redshift where the rest-frame 22$\mu$m can probe
    warm (AGN) dust, as well as PAH features.

   \subsection{Mid-Infrared Colours of the ERQs}
\begin{figure}
      \includegraphics[width=8.20cm, height=7.65cm, trim=1.3cm 0.9cm 0.3cm 0.8cm]
      {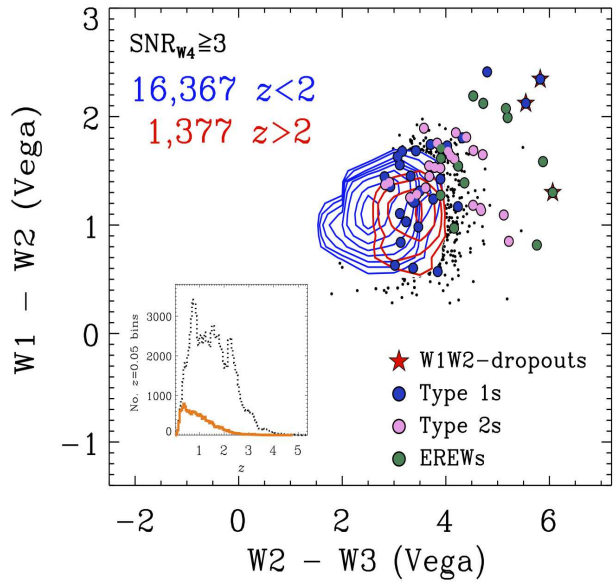}
      \centering
      \vspace{-4pt}
      \caption[]{
        The mid-infrared ``W1W2W3'' colour-colour distributions for
        objects in the SDSS and BOSS Quasar catalogues, that have a match in
        WISE and have a $W4$-band $SNR\geq3$. Objects with redshifts $z<2$ and
        $z>2$ are given by the blue and red contours, respectively, with the
        contouring level the same for the subsets of the distributions. The
        insert gives the redshift distribution for the $W4$-matched WISE
        sample which have good $W1W2W3$ detections (dashed histogram) and
        which pass the ``good W4'' criteria (solid histogram).  The large
        coloured points indicated the $r-W4>14$ quasars. We split this sample
        into three broad classes.  The EREW sample generally lies on the edge
        of the overall $W4$-matched population, with roughly half the sample
        having very red colours i.e., $W2-W3\gtrsim4.5$.  The stars signify
        objects that pass the ``W1W2-dropout'' selection of
        \citet{Eisenhardt12} (see text for details).
      }
      \label{fig:colorcolor_W1W2W3}
    \end{figure}

    The mid-infrared colour-colour distributions for the SDSS/BOSS
    quasars that are detected in WISE are shown in
    Figure~\ref{fig:colorcolor_W1W2W3}. Contours are given for the full
    ``good'' $W4$ sample, with objects that have redshifts $z<2$ and $z>2$
    given by the blue and red contours, respectively. The colours of the
    extremely red quasar sample are given by the large solid circles. We
    also split the extremely red quasar sample into three broad classes,
    `Type 1' (section~\ref{sec:Type1_egs}), `Type 2'
    (sections~\ref{sec:Type2_loz_egs} and \ref{sec:Type2_hiz_egs}) and the
    Extreme Rest-frame Equivalent Width (EREW) objects
    (section~\ref{sec:erews}).
    
    Fig.~\ref{fig:colorcolor_W1W2W3} shows that the Type 2 and EREW
    quasar populations are redder in $(W2-W3)$ than the general
    $W4$-detected population.  Upon visual inspection of the $W4$-detected
    objects with very red, $(W2-W3) \gtrsim 3.8$ colours, many were found
    to lie along the Ecliptic and did not have secure $W4$ photometry due
    to Mars, Jupiter and Saturn producing a significant number of spurious
    mid-infrared
    detections\footnote{\href{http://wise2.ipac.caltech.edu/docs/release/allsky/expsup/sec5\_3.html\#planets}{wise2.ipac.caltech.edu/docs/release/allsky/expsup/sec5\_3.html\#planets}}.
    These detections are not plotted in Fig.~\ref{fig:colorcolor_W1W2W3} and  
    does not affect any of our $r-W4>14$ extremely red quasars. 

    In $(W1-W2)$, the extremely red quasar population is perhaps
    slightly more representative of the $W4$-detected population; none of
    the red quasars are bluer than $(W1-W2)\approx0.5$. The $W1W2W3$
    colour-colour diagram has become a key diagnostic for WISE objects,
    and its power in separating different classes of objects is seen in
    \citet{Wright10, Eisenhardt12} and \citet{Yan13}. In particular, 
    \citet{Eisenhardt12} use a colour selection based on $W1W2W3$ colours
    to select ``W1W2-dropouts''. Three of our quasars satisfy this
    criterion, signified by the stars in
    Fig~\ref{fig:colorcolor_W1W2W3}. We discuss these objects in
    Section~\ref{sec:w1w2_drops}.

    \vspace{-14pt}

    \subsection{Radio Properties}
    Table~2 also gives the radio fluxes of the $r_{\rm AB} - W4_{\rm
      Vega}>14$ sample. 37 (56\%) of the full extremely red quasar sample is
    detected in the FIRST survey \citep{Becker95}, which has a flux limit
    of $\approx$1 mJy at 20 cm, with 22 out of 28 objects at $z<1.1$
    (79\%) being detected. The magnitude of radio flux at 20cm ranges from
    0.55 to 960 mJy.
    
    While the radio loudness is typically defined by the ratio of the 5 GHz radio to 4400\AA\ flux densities \citep[e.g.,][]{Kellermann89, Padovani03}, with the advent of the FIRST and SDSS surveys, much of the recent analysis of radio loudness has been done in terms of the 20cm to $i$-band ratio, $\mathcal{R} = f_{\rm radio} / f_{\rm optical}$, e.g., \citet{Ivezic02}, \citet{White07radio} and \citet{Kratzer15}. Typically no $k$-correction is applied as the mean spectral index in the radio and optical are similar (though this assumption may not hold true for these extremely red, and unusual objects). While there is no true gap in the distribution of $\mathcal{R}$, at $\mathcal{R}$$\sim$10 there is a subtle change in the distribution \citep[e.g.,][]{White07, Kratzer15} and it is common to refer to objects with $\mathcal{R}\geq10$ as ``radio-loud''.  The value of $\mathcal{R}$ can be calculated straight from Table 2 for each of our objects, and we find that all 37 of our radio-detected objects are nominally radio-loud. However, without knowing the extent of reddening from dust in our sources, these values should be considered as an upper limit to the true value of $\mathcal{R}$. We also find that all of the 37 radio-detected objects are unresolved in FIRST.
    
    A total of 30 objects (46\%) in our sample have a radio target selection flag set indicating that these objects were detected in the FIRST survey and passed the SDSS/BOSS quasar selection to be observed as quasar targets. Of these 30 objects, 10 are from SDSS and 20 from BOSS. 17 objects (26\%) are selected for spectroscopy {\it solely} due to their radio properties, all of which are from BOSS.
    
    \citet{Yan13} discovered ULIRGs at $z\sim2$ with extremely red colours, which suggests that star-formation will lead to an enhancement of the $W4$ emission we detect from our sample. While the sample definition of $r-W4>14$ selects quasars that are red due to dust obscuration (i.e., the targeted obscured quasar population), there is the possibility it will also select systems with particularly bright IR emission due to star formation. The large fraction of radio-detected quasars in the resulting sample, and the known relations between radio and IR emission from star formation, suggest that this may be an aspect of the extremely red quasar selection.
    
    \begin{figure}
      \includegraphics[width=8.60cm, height=7.30cm, trim=0.0cm 10.5cm 10.5cm 1.0cm, clip]
      {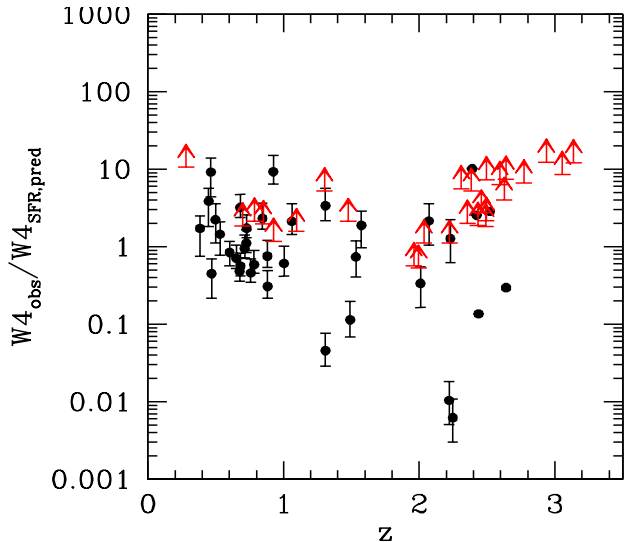}
      \centering
      \vspace{-8pt}
      \caption[]{The ratio between the observed $W4$ flux of our sample 
        and the $W4$ flux that would be expected given the observed 
        radio luminosities and the conversions between radio and star
        formation rates reported in \citet{Bell03_SFRs}. 
        A ratio of $>1$ means that the $W4$ emission is likely AGN dominated.
        For objects undetected in the radio (FIRST) survey, we only calculate an
        upper limit on the star formation rate and thus a lower limit on the
        observed-to-predicted flux ratio, given by the red arrows.}
      \label{fig:qir}
      \vspace{-12pt}
    \end{figure}
    
    To investigate this further, we calculate rest-frame 1.4 GHz radio
    luminosity from the observed fluxes using a radio spectral index of
    $\alpha=-0.7$. In the first instance, we assume that all of the radio
    emission is due to star formation, which will result in upper limits
    on the SFR derived from the radio and MIR. We use the conversions
    between radio and star formation rates reported in \citet{Bell03_SFRs}
    to calculate the necessary SFRs and total IR luminosity of star
    formation $L_{\rm SFR,\ 8-1000\mu m}$. Then we scale seven
    star-formation templates from \citet{Kirkpatrick12} and
    \citet{Mullaney11} to match the SFRs necessary to explain the observed
    radio emission. We then redshift them to the redshift of the target,
    convolved with the $W4$ filter from \citet{Jarrett11} and calculate
    the predicted IR flux. Figure~\ref{fig:qir} shows the ratio between
    the observed and the predicted $W4$ fluxes; those are shown as black
    points, where the error bars reflect the spread amongst templates (and
    therefore provide a measure of the systematic error). For objects
    undetected in the radio, we can only calculate an upper limit on the
    SFR rate and thus a lower limit on the observed-to-predicted flux
    ratio; those are shown as red upward arrows, and those are
    conservative (i.e., the lowest lower limit of the seven templates). A
    ratio of $> 1$ means the MIR emission is likely AGN dominated, while
    ratios $< 1$ mean the predicted SFR is less than expected, given the
    radio flux.
    
    There are 37 objects detected in the radio; if all the radio flux
    is associated with star-formation, which is unlikely, then for these
    objects the median luminosity due to star formation would need to be
    $L_{\rm 8-1000\mu m}=10^{13.6}L_{\odot}$ in order to explain the
    median radio luminosity of $\nu L_{\nu}[{\rm 1.4GHz}]=10^{41.6}$ erg
    sec$^{-1}$. In other words, these objects would need to be forming
    stars at the median rate of $\approx$7700 $M_{\odot}$ yr$^{-1}$ in
    order to explain the observed radio emission, and this rate is
    $\sim$4-10$\times$ higher than the e.g. 850$\mu$m-selected
    submillimeter galaxy (SMG) population at redshifts $z\sim2-3$
    \citep[][]{Wardlow13, Fu13, Casey14}. However, this SFR is an upper
    limit, as the AGN is almost certainly contributing significant flux in
    all objects. It is more likely that both the radio emission and the
    $W4$ emission in these objects are due to the AGN and star formation,
    and thus the SFRs are more in-line with those of SMGs and
    WISE-selected luminous infrared galaxies \citep{Wu12_w1w2drops,
      Jones14, Jones15, Tsai15}.

    The two points with the lowest $W4_{\rm obs}/W4_{\rm SFR,
      predicted}$ ratios are potentially beamed sources with an apparent
    extremely high radio luminosity $\nu L_{\nu}[{\rm 1.4GHz}]\simeq
    10^{44.5}$ erg s$^{-1}$ which has nothing to do with star formation in
    the host galaxy (though our data are insufficient to determine the
    origin of the $W4$ emission in these sources). For the less extreme
    sources at $z<1$, we find that the required median SFR necessary to
    explain the observed radio and W4 emission is 2180 $M_{\odot}$
    yr$^{-1}$, which is still very high, but could potentially be the
    result of an e.g. (major) merger.  While we do not have direct
    observations (e.g., far-infrared) which would rule out such star
    formation rates, various authors find it unlikely that star formation
    dominates mid-infrared and radio emission of luminous Type 2 quasars
    at $z<1$ \citep[e.g.,][in prep.]{Harrison14, Zakamska_Greene14,
      Zakamska15}.
    
    The remaining objects are undetected in the radio and therefore we
    can place only upper limits on star formation rates. For this
    population we find the median lower limit $W4_{\rm obs}/W4_{\rm SFR,
      predicted}$ of 2.4, meaning that the star formation is inadequate by
    at least this factor in explaining the observed W4 flux. Thus we
    conclude that the $W4$ fluxes in these objects must be dominated by
    the AGN as well.  However, it seems that ultimately, we currently do
    not have enough data to determine whether the radio emisson is due to
    AGN or star formation. New FIR/submm data is required, with further
    SED decomposition, and indeed this is where future analysis will be
    concentrated (Hamann et al., 2015, in prep.; Zakamska et al., 2015, in
    prep.).

\section{Classifications from optical Spectroscopy}
\label{sec:spectral_egs}
The spectra of the extremely red quasar sample are publicly available
at the SDSS-III Science Archive
Server\footnote{\href{http://dr10.sdss3.org}{http://dr10.sdss3.org}}. We
broadly classify the objects that satisfy the $r-W4>14$ selection
into: {\it (i):} unobscured (but reddened) Type~1 quasars possessing
broad lines (bitwise value 1 in Table~3); {\it (ii):} objects that
show evidence for being narrow-line Type 2 quasars from their optical
spectra (bitwise value 2 in Table~3); {\it (iii):} objects that
suggest ongoing, or recently ceased star formation (bitwise values 4
and 8); {\it (iv):} radio-detected quasars (in the FIRST survey, bitwise value
16); {\it (v):} quasars with blue-shifted absorption features (bitwise
value 32); {\it (vi):} objects that are not detected in the WISE
$W1/2$-bands (bitwise 64) and {\it (vii):} a class of Type 1 sources
with extreme rest equivalent widths in their UV broad lines, namely,
REW$>150$\AA\ in either \civ or \mgii (bitwise 128). These
classification groups are not mutually exclusive; several objects are
assigned to multiple groups. However, we class all of the sources as
either Type 1 (37/65; 57\%) or Type 2 (28/65; 43\%). These sources are
indicated by the bitwise values in the last column of Table~2 and are
described in Table~\ref{tab:ERQ_bitmask}. Example spectra are
presented in what follows.

\begin{figure*}  
  \includegraphics[width=16.0cm, height=21.2cm, trim=0mm 00mm 0mm 5mm, clip]
  {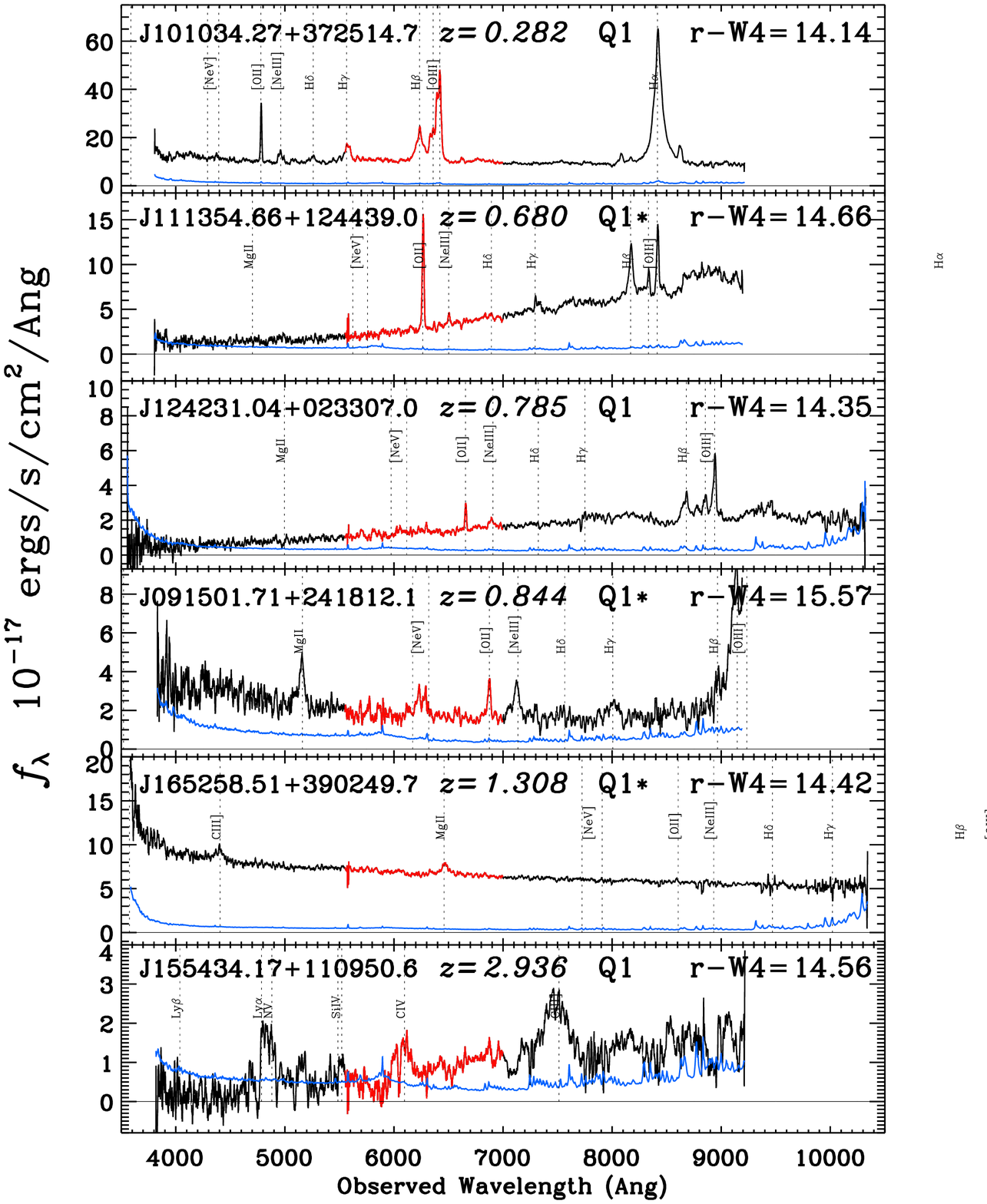}
  \centering
  \vspace{-16pt}
  \caption[]{Examples of extremely red quasars classified as Type 1 quasars based
    on their optical spectroscopy. The SDSS name, redshift, classification
    and $r-W4$ colour are given in each panel. The red part of the spectrum
    marks the contribution to the observed $r$-band flux, while the blue
    line is the estimated error per pixel \citep[][]{KGLee13}. The spectra are 
    smoothed using an 8-pixel boxcar average. Prominent emission
    lines often seen in AGN spectra;
    Ly$\beta$ 1025.72,
    Ly$\alpha$ 1215.67, 
    \nv 1240.14, 
    \SiIV 1393.76, 1402.77, 
    \civ 1548.19, 1550.76 
    \ciii 1908.73, 
    \mgii 2798.75 
    \nev 3346.82, 3426.84 
    \oii 3728.48 
    \neiii 3869.85, 
    \hdelta 4102.89, 
    \hgamma 4341.68, 
    \hbeta 4862.68, 
    \oiii 4960.30, 5008.24 and 
    \halpha 6564.61 \AA ,  
    are marked
    by the vertical dashed lines, where we use vacuum wavelengths for
    identifying emission lines. The feature at 5575\AA\ in the lower two
    panels is the \oi bright sky line. Spectra that cut off at 3800 \AA\ and
    9200 \AA\ are SDSS spectra, while BOSS spectra extend to 3600 \AA\ and
    10,400 \AA . Objects with starred classifications were targeted due to
    their radio properties. Note that the $y$-axis scales are different in each panel.}
  \label{fig:Type1_egs}
\end{figure*}

    \subsection{Type 1 Quasars}
    \label{sec:Type1_egs}
    Figure~\ref{fig:Type1_egs} presents the spectra of six examples of
    extremely red quasars, classified as Type 1 based on broad emission 
    lines seen in their optical spectroscopy. Where measured, the mean line widths
    are FWHM(Mg\,{\sc ii}) $\sim6440$ km s$^{-1}$ and FWHM(C\,{\sc iv})
    $\sim4900$ km s$^{-1}$. Figure~\ref{fig:Type1_egs} lists the SDSS
    name, redshift, classification (from Table~\ref{tab:ERQ_bitmask}) and
    $r-W4$ colour of each object. 
    
    The spectra in Figure~\ref{fig:Type1_egs} are heterogeneous; some
    show blue continua, while some are red. The first object (with the
    lowest redshift in our sample) has unusually broad [O\,{\sc iii}] 
    with non-Gaussian structure \citep[][see also
    section~\ref{sec:starburst_egs}]{Liu10}. Many of the examples of Type
    1 quasars from our selection appear to be significantly reddened
    \citep{Richards03} often with a host galaxy component. 

    \citet{Glikman13} identify dust-reddened quasars by matching the FIRST
    radio catalogue to the UKIRT Infrared Deep Sky Survey \citep[UKIDSS;
    ][]{Lawrence07}. This study identified 14 reddened quasars with $E(B -
    V ) > 0.1$, including three at $z > 2$. However, \citet{Glikman13}
    find no heavily reddened $(E (B - V )\gtrsim 0.5)$ quasars at high
    redshifts $(z > 2)$. The object ULASJ1234+0907 $(z = 2.503)$
    discovered by \citet{Banerji14} is currently the reddest broad-line
    Type 1 quasar known, with $(i - K)_{\rm AB} > 7.1$. Since all our
    extremely red quasars have $i \lesssim 22$ mag and a range of $1.62 <
    (i - K)_{\rm AB} < 4.73$, none of our objects have optical-to-near
    infrared colours as red as ULASJ1234+0907.
    
    We note the object J165258+390249 has a relatively blue continuum
    and is clearly not suffering from dust obscuration. This object may be
    optically variable, and at the time of the optical and IR photometry
    had a large enough $r-W4$ to satisfy our colour cut. We also note it
    is a radio source, and has features consistent with an optical blazar
    spectrum. This further highlights the importance of understanding the
    radio properties (and potentially the synchrotron radiation
    contribution) of these quasars and how it affects the IR and the
    resulting sample of objects.

    \begin{figure*}  
      \includegraphics[width=16.0cm, height=21.6cm, trim=0mm 00mm 0mm 5mm, clip]
      {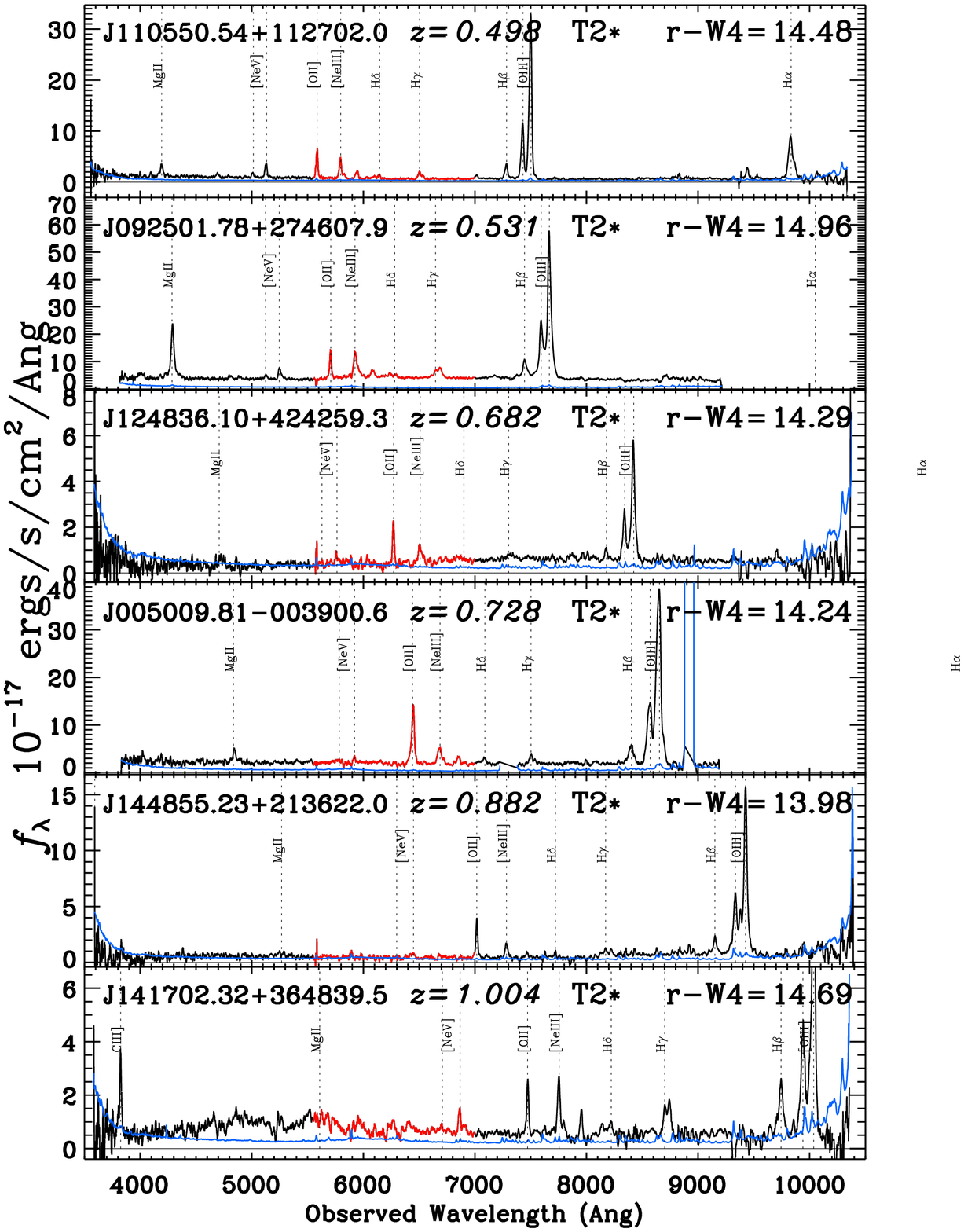}
      \centering
      \vspace{-8pt}
      \caption[]{Same as for Fig~\ref{fig:Type1_egs}, except 
        showing examples of extremely red quasars classified as
        Type 2 quasars based on their optical spectroscopy, 
        at $z\lesssim1.1$. }
      \label{fig:Type2_loz_egs}
    \end{figure*}
    \subsection{Type 2 Quasar candidates at low redshift}
    \label{sec:Type2_loz_egs}
    We define Type 2 quasars as objects with narrow permitted emission
    lines that do not have an underlying broad component, and with line
    ratios characteristic of non-stellar ionizing radiation
    \citep{Zakamska03, Reyes08}. Specifically, in Type 2 quasars the
    broad-line region is completely blocked from the observer by strong
    obscuration ($A_{V}>10$), and nuclear activity is inferred indirectly
    via the emission line ratios of ionized gas which are characteristic
    of photo-ionization by the quasar \citep{BPT, Veilleux87}. At low
    redshifts, where H$\beta$ and \oiii are accessible in the optical
    spectrum, we can define Type 2 objects as those in which the kinematic
    structure of the permitted emission lines is similar to that of the
    forbidden lines; the permitted lines should not display an additional
    broad (FWHM$\ga 2000$ km sec$^{-1}$) component. We also require
    \oiii/H$\beta>3$.
 
    Figure~\ref{fig:Type2_loz_egs} presents examples of Type 2 quasar
    candidates at $z\lesssim1$. No line-fitting has been performed, but
    all objects in this class are seen by visual inspection to have
    similar same kinematics in the permitted H$\beta$ and forbidden \oiii
    lines. In the extremely red quasar sample, the quasar nature of the
    Type 2 candidates is not in doubt: the typical values of
    \oiii/H$\beta$ are 10 or greater, and high-ionization species such as
    [NeV] are detected in most cases. However, H$\beta$ can be weak and
    there may be broad components in some H$\alpha$ lines (Zakamska et
    al. in prep) which are redshifted out of the BOSS wavelength coverage. 

    Just over half (60\%) of our Type 2s are at $z<1$. This is due to
    the fact that we are using the spectroscopic quasar catalogues as our
    initial sample, before matching to the WISE photometry. Thus, from the
    requirement of having a relatively bright detection in the optical
    (cf. the Brand et al. studies) there will be a selection bias towards
    lower redshift objects.
    
    Among the 28 objects with $z<1.1$, 20 (71\%) are classified as
    Type 2 candidates from their optical spectra; see
    Figure~\ref{fig:Type2_loz_egs}. This high fraction of Type 2 objects
    is completely uncharacteristic of the parent sample of the SDSS/BOSS
    quasars, making it clear that the selection based on high
    infrared-to-optical colours can indeed recover obscured quasars. All
    seven Type 2 quasars in our sample that were observed before July 2006
    are found in the catalogue of \citet{Reyes08}.

    \begin{figure*}  
      \includegraphics[width=16.0cm, height=21.6cm, trim=0mm 00mm 0mm 5mm, clip]
      {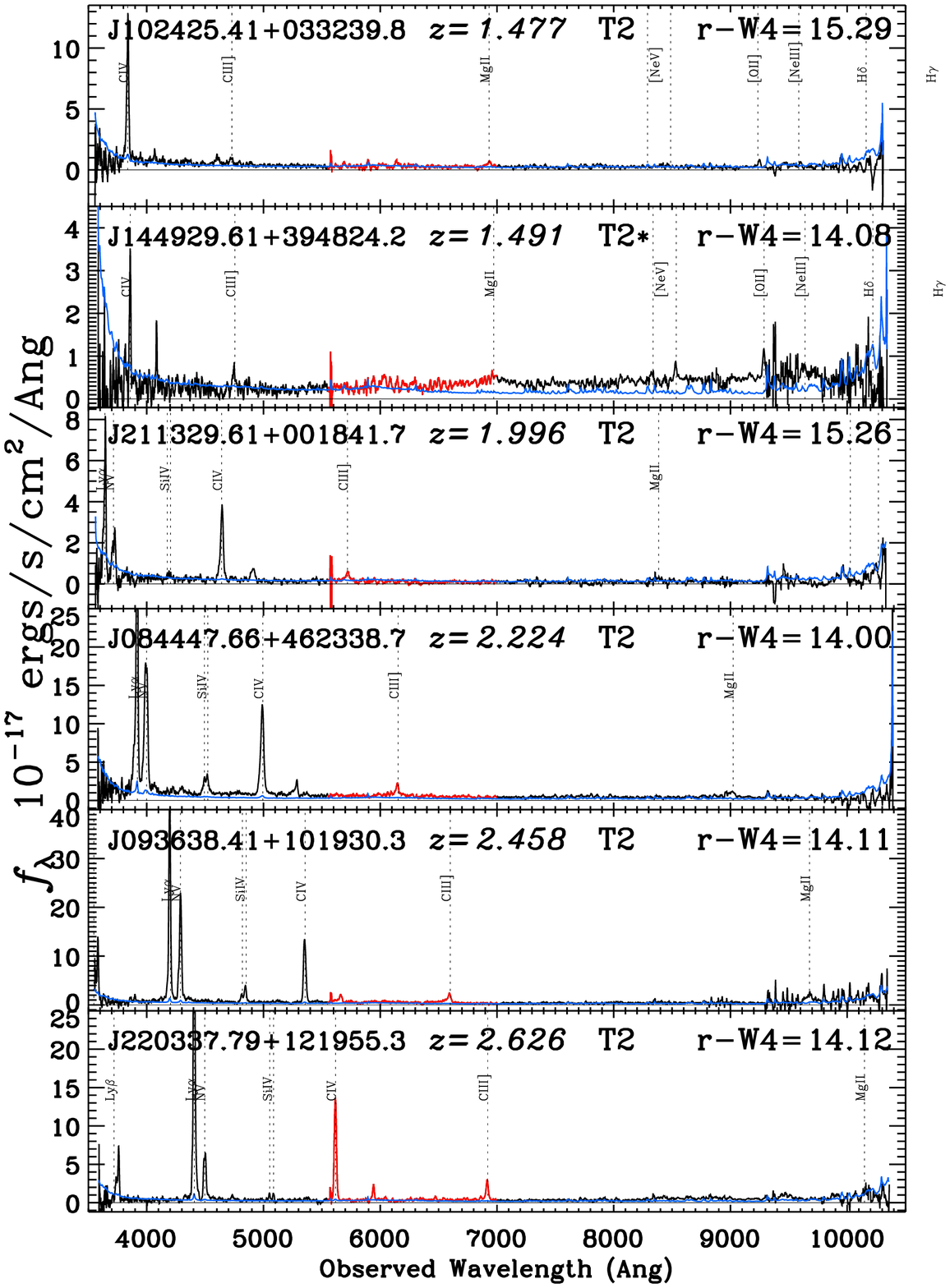}
      \centering
      \vspace{-8pt}
      \caption[]{Same as for Fig~\ref{fig:Type1_egs}, except 
        showing examples of extremely red quasars classified as
        Type 2 quasars based on their optical spectroscopy, 
        at $z\gtrsim1.5$. }
      \label{fig:Type2_hiz_egs} 
    \end{figure*}
    \subsection{Type 2 Quasar candidates at high redshift}
    \label{sec:Type2_hiz_egs}
    We find objects that have narrow emission lines, weak continua and
    satisfy the definition of Type 2 candidates, at high redshift
    ($z>1.5$) as well; see Figure~\ref{fig:Type2_hiz_egs}. The
    classifications are performed by visual inspection of the optical
    spectra and all objects in this class with $z>1.5$ have FWHM(C\,{\sc
      iv}) $\lesssim2000$ km s$^{-1}$.

    These objects have similar optical spectra to those described in
    \citet{Yan13} and \citet{Alexandroff13}, although there are no objects
    in common between our sample and that presented in
    \citet{Alexandroff13}. Those authors present a sample of 145 candidate
    Type 2 quasars from BOSS at redshifts between 2 and 4.3, using data
    from DR9 \citep{DR9}. These objects are characterized by weak
    continuum in the rest-frame ultraviolet with a typical continuum
    magnitude of $i\approx22$ (i.e., $M_{i}(z=2.5)\approx-25$) and strong
    lines of \civ and Ly$\alpha$, with FWHM $\leq 2000 \kms$. Typical ($\sim
    L^{*}$) galaxies are $\geq 1$ magnitude fainter at these redshifts
    \citep{Marchesini07, Marchesini12}, suggesting host-galaxy light is
    not sufficient to explain the continuum luminosity and some
    level of AGN light is getting through/around the absorber.
    
    The reddest quasar in \citet{Alexandroff13} has $r-W4=13.30$. In
    general, our selection of Type 2 candidates is not as strict as that
    of \citet{Alexandroff13}, and we find no overlap between the sample
    here and the DR9-based sample in \citet{Alexandroff13}. While this
    class of object shows only narrow emission lines in the rest-frame UV
    spectra, \citet{Greene14} obtained near-infrared spectroscopy for a
    subset of the \citet{Alexandroff13} sample, demonstrating that the
    H$\alpha$ emission line consistently requires a broad component. This
    result implies that the typical extinction in these objects is
    $A_{V}\lesssim$ a few, sufficient to block the rest-frame UV continuum
    and broad lines but not the rest-frame optical - consistent with the
    (relatively) bright continuum.

    \begin{figure*}  
      \includegraphics[width=16.0cm, height=21.6cm, trim=0mm 00mm 0mm 5mm, clip]
      {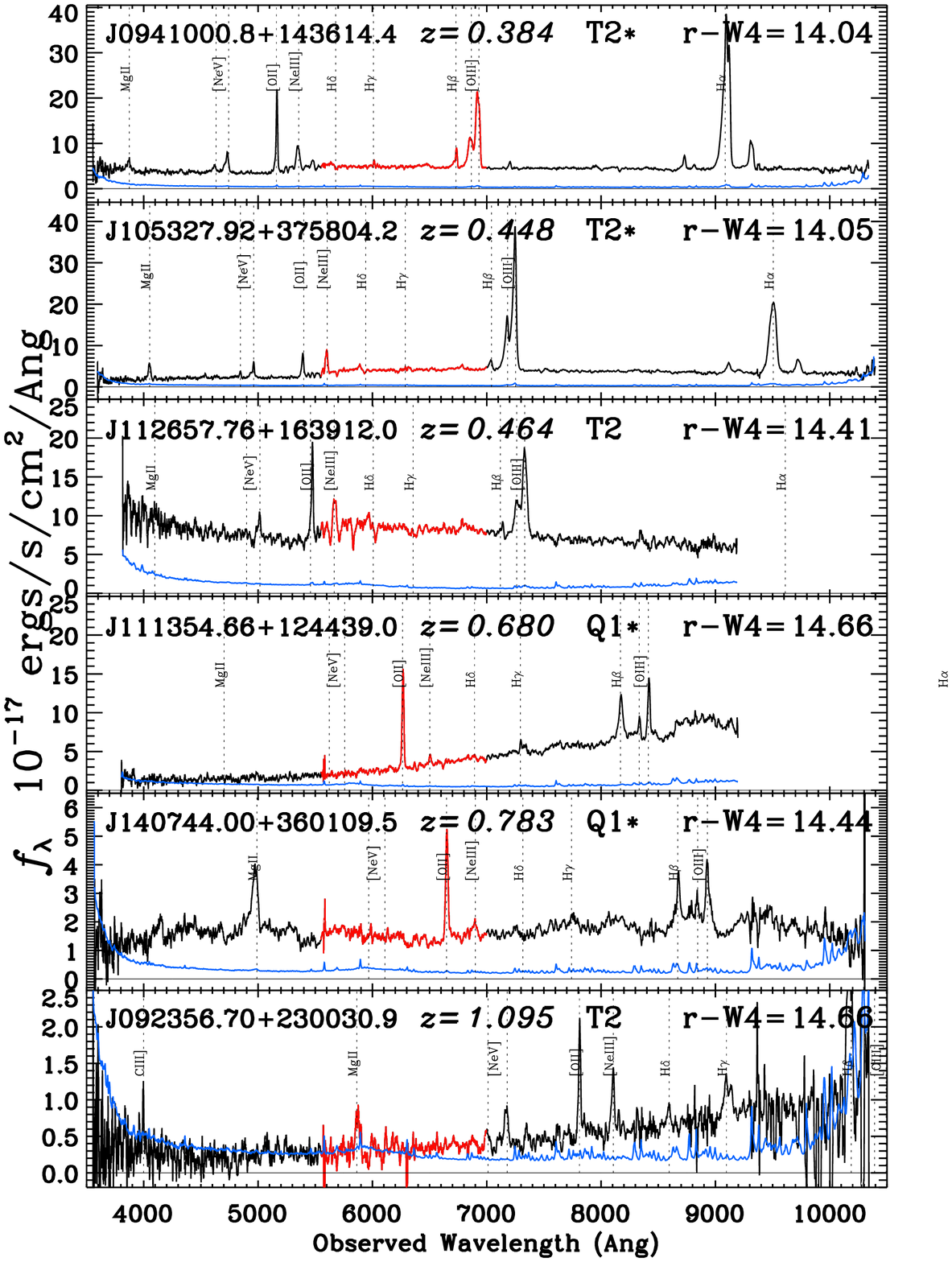}
      \centering
      \vspace{-8pt}
      \caption[]{Same as for Fig~\ref{fig:Type1_egs}, except 
        showing examples of extremely red quasars that display signatures of 
        an ongoing, or post, star-bursting phase. 
        In particular we identify the Balmer edge in 
        J105327.92+375804.2 and J112657.76+163912.0.} 
      \label{fig:starburst_egs} 
    \end{figure*}
    \subsection{Objects with starburst and post-starburst signatures}
    \label{sec:starburst_egs}
    Our sample includes quasars that have starburst or post-starburst
    signatures in their optical spectra. Examples are given in
    Figure~\ref{fig:starburst_egs}.
    
    The [OII]/[OIII] ratio can be used as a starburst indicator in AGN
    and galaxies \citep[e.g.,][]{Kauffmann03, Groves06}. Objects from our
    sample are placed into the starburst class if they exhibit evidence of
    obvious \oii emission \citep{Kennicutt98, Kewley02, Kewley04, Ho05,
      Moustakas06, Mostek12}, though we acknowledge that [OII] emission can
    also be associated with the narrow-line region of the AGN.
    
    The post-starburst objects are objects which have either a Balmer
    break, and/or an obvious Balmer series, originating from an A-star
    population that is not dominated by blue light from OB-stellar
    populations. Quasars with post-starbursting signatures have been
    identified previously, \citep[e.g.,][]{Brotherton99, Brotherton02,
      Cales13, Cales14AAS} and \citet{Matsuoka15} derive physical properties
    from optical spectra while \citet{Wei13} study the MIR spectral
    properties of post-starburst quasars.

    Both post-starburst objects in our sample are at relatively low
    redshift, and not all the host galaxy light is likely to be captured
    by the 2'' spectroscopic fiber. However, the quasar
    J112657.76+163912.0 (hereafter J1126+1639) has a pronounced Balmer
    break and we estimate that the fraction of e.g A-star host galaxy flux
    required is at least that of the AGN. J1126+1639 may have other
    optical objects within the $W4$-beam, as we mentioned in
    Section~\ref{sec:erq_criteria}. J1126+1639 appears to be at the
    center of a group/small cluster of other optically red
    galaxies. However, the other group members are {\it blue} in $(W1-W2)$
    colour, suggesting that these objects are not bright in $W4$. Our
    likely candidate appears disturbed in the SDSS optical image,
    suggesting that it hosts an ongoing merger.

    \begin{figure}  
      \includegraphics[width=8.6cm, height=8.2cm, trim=5mm 10mm 0mm 5mm, clip]
      {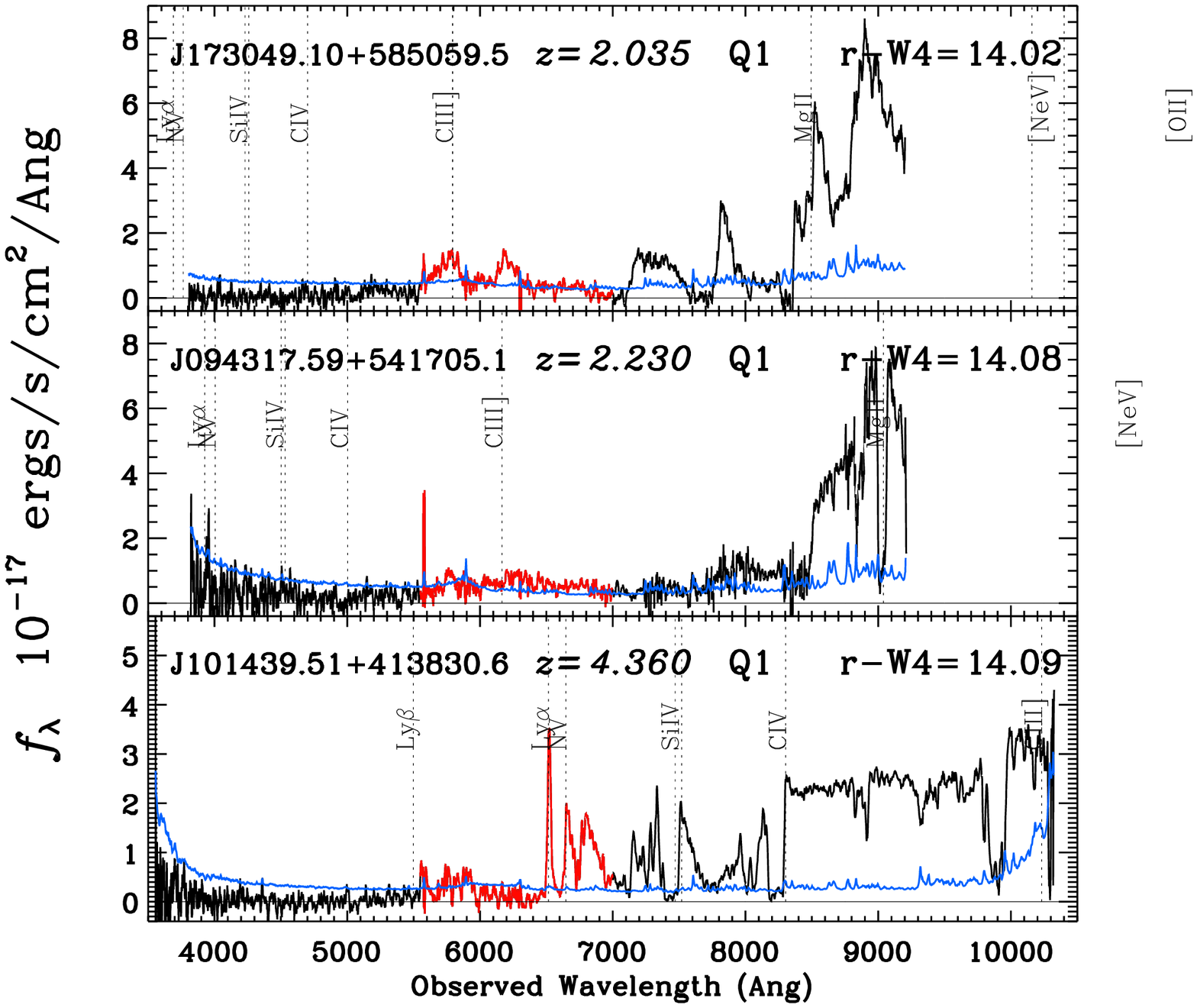}
      \centering
      \vspace{-8pt}
      \caption[]{Same as for Fig~\ref{fig:Type1_egs}, except showing examples of
        the Broad Absorption Line quasars. These three objects are  
        examples of the rare subclass of FeLoBALs. 
        At $z=4.36$, J101439.51+413830.6 is
        the highest redshift object in our full sample.}
      \label{fig:BAL_egs} 
    \end{figure}
    \subsection{Broad Absorption Line Quasars}\label{sec:BALs}
    Our $r-W4>14$ quasar selection also finds broad absorption line
    (BAL) objects (Figure~\ref{fig:BAL_egs}). Although we do not use it
    for our classifications, the \civ balnicity index \citep[BI;
    ][]{Weymann91} for our extremely red BAL quasars as reported in the
    DR10Q catalogue is generally $\log_{10} ({\rm BI}_{\rm C IV}) \gtrsim
    3$. In particular, we find examples of BALs that exhibit absorption
    from low-ionization species, i.e. LoBALs, and objects that have \feii
    absorption, i.e. FeLoBALs. At $z=4.36$, J101439.51+413830.6 is the
    highest redshift object in the full sample, but due to this high
    redshift, any \feii absorption will be outside the optical wavelength
    coverage of the SDSS/BOSS spectra.
    
    It has been known for some time that BALs can be redder (in
    optical-to-near infrared colour) than the general quasar population
    e.g., \citet{Hall97}. \citet{Hall02} report on a number of
    heavily-reddened, extreme BAL quasars discovered in SDSS, while
    \citet{Trump06}, \citet{Gibson09} and \citet{Allen11} explore this
    reddening-BAL relation. However, we note that the presence of the BAL
    objects in our sample is not due just to dust reddening; in these
    objects, the BAL troughs remove a large fraction of the continuum that
    would otherwise contribute to the $r$-band flux, which makes the
    $r-W4$ colour even redder. We find FeLoBALs in our sample not because
    they have red continua, but because the BALs have wiped out the
    $r$-band flux; in this sense these are in the sample for different
    reasons than for the other very red quasars.

    \begin{figure}  
      \includegraphics[width=8.8cm, height=8.2cm, trim=16mm 10mm 0mm 5mm, clip]
      {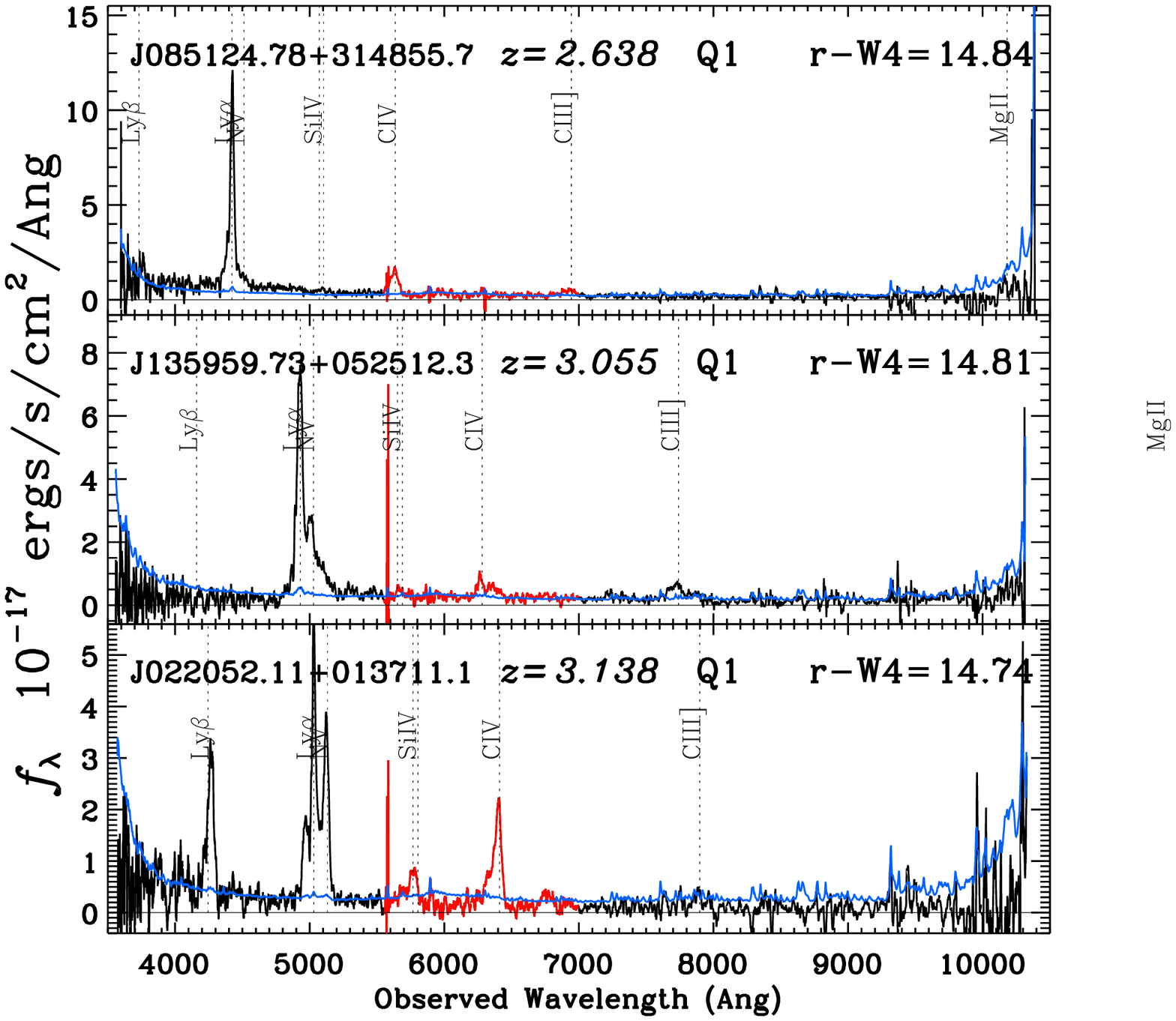}
      \centering
      \vspace{-8pt}
      \caption[]{Same as for Fig~\ref{fig:Type1_egs}, except showing quasars
        that are ``W1W2-dropouts''. The feature at 5575\AA\ in the lower two
        panels is the \oi bright sky line.}
      \label{fig:W1W2drops_egs} 
    \end{figure}
    \subsection{W1W2-dropout Quasars}\label{sec:w1w2_drops}
    Objects that are selected to be bright at 12 or 22$\mu$m, but are 
    undetected by WISE at 3.4 and 4.6$\mu$m are
    ``W1W2-dropouts''. \citet{Eisenhardt12} find $\sim$1000 such objects
    over the whole sky, including the $z=2.452$ source WISE
    J181417.29+341224.9, a hyper-luminous infrared galaxy (HyLIRG) with
    $L_{\rm IR} > 10^{13} L_{\odot}$. WISE 1814+3412 has a $W4$ flux
    density of $14.38\pm0.87$mJy; a 350$\mu$m detection
    \citep{Wu12_w1w2drops} implies a minimum bolometric luminosity of
    $3.7\times10^{13} L_{\odot}$ suggesting that the W1W2-dropouts are
    extreme cases of luminous, hot (60-120K) dust-obscured galaxies
    possibly representing a short evolutionary phase during galaxy merging
    and evolution. Due to these dust temperatures, the W1W2-dropouts are
    also known as hot, dust-obscured galaxies (``Hot DOGs'').
    
    Three of our 65 objects ($\sim$5\%) satisfy the W1W2-dropout
    selection criteria. Their optical spectra are shown in
    Figure~\ref{fig:W1W2drops_egs}. These three quasars were all
    discovered by BOSS, have $r$-band fluxes of $\approx 5-8$ $\mu$Jy
    (cf., $2.25\pm0.11$ $\mu$Jy for WISE 1814+3412) and redshifts
    $z=2.638-3.138$, placing them in the HyLIRG regime.

   \begin{figure*}
      \includegraphics[width=6.3in, height=8.5in]
      {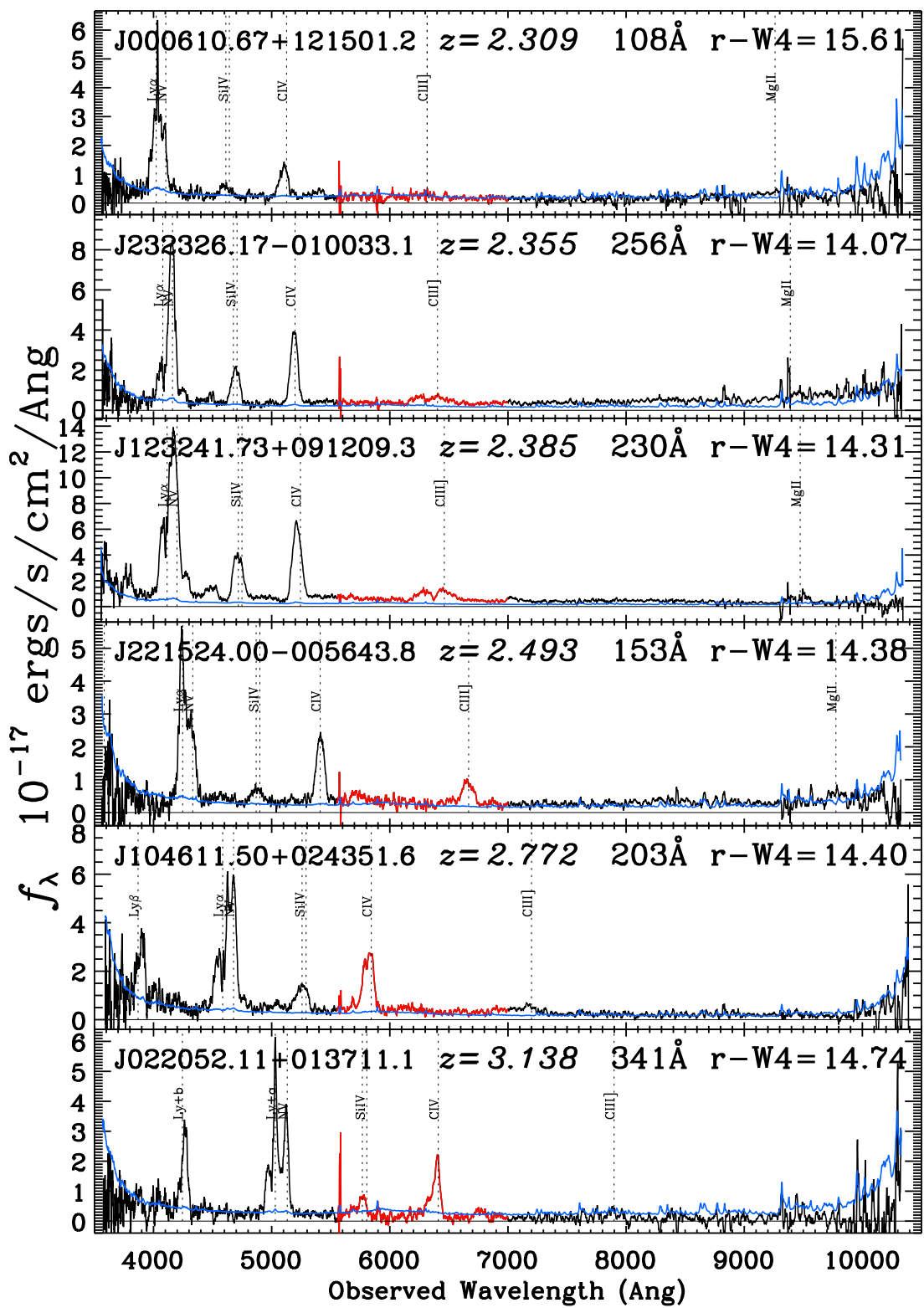}
      \centering
      \caption[]{Same as for Fig~\ref{fig:Type1_egs}, except showing examples of
        the new set of ``Extreme Rest-frame Equivalent Width'' (EREW)
        objects. These objects possess large REWs of e.g. the \civ line
        (values given in Table~2), unusual profiles of the Ly$\alpha$ to \nv 
        complex (and the ratio of Ly$\alpha$:NV) and flat continua in all
        examples. The quasar J022052.11+013711.1 (bottom panel) is also a
        W1W2-dropout (Fig.~\ref{fig:W1W2drops_egs}).}
      \label{fig:EREWs_egs}
    \end{figure*}
    \subsection{Extreme Equivalent Width Objects}
    \label{sec:erews}
    The objects that we present in Figure~\ref{fig:EREWs_egs} all have
    \civ FWHM $>2000 \kms$ (Table 2) and are thus classified as Type 1
    quasars. These six examples are a set of objects at $z\simeq 2-3$
    which are characterized by extreme rest-frame equivalent widths
    (EREWs) of $\gtrsim150$\AA\ (the measured REWs from the \civ line are
    given in Fig.~\ref{fig:EREWs_egs}). Four sources have REW(C\,{\sc iv})
    $>200$\AA.  For comparison, typical quasars have REW(C\,{\sc
      iv})$\approx$25-50\AA. Some of the extreme REW sources also have
    unusual line properties, including high N\,{\sc v}/Ly$\alpha$,
    enhanced Si\,{\sc iv}+O\,{\sc iv}] and weak He\,{\sc ii} and C\,{\sc
      iii}]. Some also have strong \lyb and unusual \lya profiles.
    
    To quantify how unusual these objects are, we use emission line
    data from the DR10Q catalogue to select sources that have: $2.0 < z <
    3.4$, the visual inspection flag for BALs set to 0 (i.e. no BALs),
    FWHM(C\,{\sc iv}) $>2000$ km s$^{-1}$, and the ratio of the \civ
    rest-frame equivalent width to its uncertainty $>3$. A total of 98807
    quasars satisfy these criteria and in the cumulative distribution,
    5.8\% of objects have REW(C\,{\sc iv}) $>100$\AA\ with 1.3\% having
    REW(C\,{\sc iv}) $>150$\AA. This is to be compared to 55\% (45\%) of
    the non-BAL Type 1s with measurable CIV in our sample having
    REW(C\,{\sc iv}) $>100$ (150) \AA.
    
    The quasar J1535+0903 (see Appendix~\ref{sec:J1535}) may also be a
    member of this class. We measure REW(C\,{\sc iv})=136 \AA, and
    REW(Mg\,{\sc ii}) $\sim$280 \AA, the precise value depending on
    uncertain continuum placement. Like the N\,{\sc v}/\lya $>1$ objects,
    J1535+0903 has very strong \aliii $\lambda$1860.

    The large REWs in these objects might be caused by suppressed
    continuum emission analogous to Type 2 quasars in the Unified
    Model. However, the large line widths \citep[at least compared to
    typical narrow line regions;][]{Liu13b} and in some cases very high
    electron densities (Section~\ref{sec:J1535}; Hamann et al., in prep.)
    suggest that the line-emitting regions are close to the central
    continuum source, as in typical broad line regions. It therefore seems
    difficult to have dust obscuring the continuum source without also
    obscuring the broad emission lines. We investigate these issues
    further in Hamann et al. (in prep.).

    \begin{figure}
      \centering 
      \includegraphics[width=8.25cm, height=7.25cm]
      {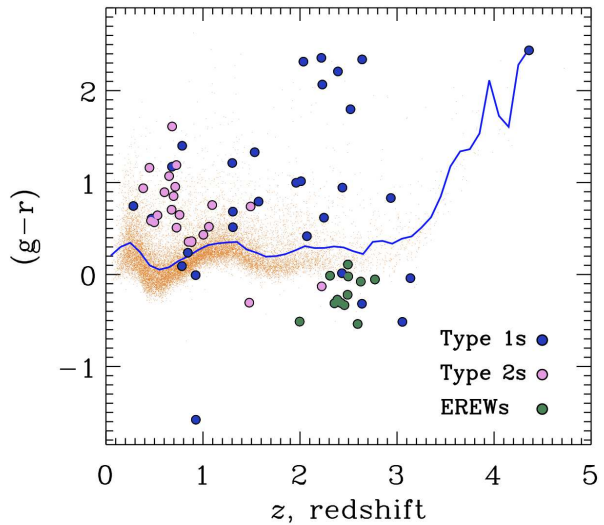}
      \vspace{-6pt}
      \caption{
        The optical $(g-r)$ colour of the $W4$-matched quasars (orange
        points) as a function of redshift. The mean colour of the $W4$-matched
        sample is given by the solid blue line.  At $z>1.9$ the colour
        distribution of our extremely red quasars is impressively broad,
        demonstrating the heterogeneity of our sample.  The EREW population is
        {\it bluer} than the overall population in $(g-r)$ at $z\approx2.5$.
        All optical magnitudes are AB.
      }
      \label{fig:ERQs_ugriz_colors_vs_redshift}
    \end{figure}

    \begin{figure}
      \centering 
      \includegraphics[width=8.5cm, height=10.5cm, trim=1.2cm 0.9cm 0.6cm 0.8cm]
      {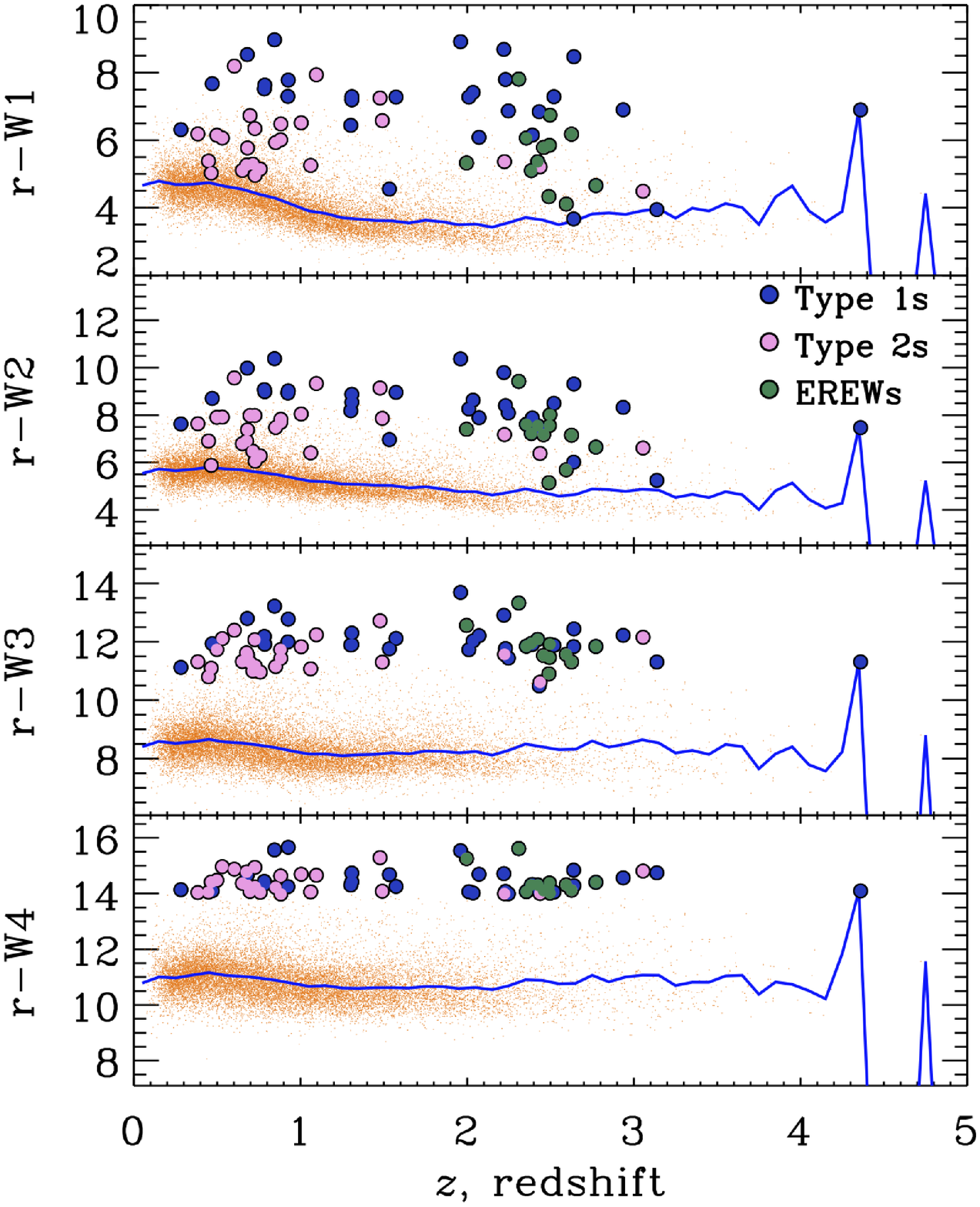}
      \vspace{-14pt}
      \caption{
        The optical-midinfrared colour for our $W4$-matched sample 
        (orange points), with $r-W1$, $r-W2$, 
        $r-W3$ and $r-W4$ shown from top to bottom. 
        The mean colour of the $W4$-matched sample is given by the solid blue line.
        As the optical-MIR wavelength arm extends from 
        $W1$ (3.4$\mu$m) to $W4$ (22$\mu$m) the selected population 
        moves further away from the mean $W4$-matched sample. 
        All optical magnitudes are AB, all WISE magnitudes are Vega.
      }
      \label{fig:ERQs_rW1234_colors_vs_redshift}
    \end{figure}

 
\section{Colours and selection of the Extremely Red Quasars}
\label{sec:spectral_selection}
The optical SDSS/BOSS spectra of the extremely red quasars have been
presented in Section 4. The heterogeneity of the optical spectra is
immediately obvious. Upon visual inspection of the optical spectra, we
have classified the extremely red quasar sample into various groups
given in Tables~2 and \ref{tab:ERQ_bitmask}. In this section, we
explore the colours and selection of the extremely red quasars, in
order to understand the relationship of the sub-classes to each other,
and with respect to the general quasar population.

    \subsection{Colours of the extremely red quasar groups}
    Figure~\ref{fig:ERQs_ugriz_colors_vs_redshift} presents the
    optical $(g-r)$ colours for the full $W4$ matched sample as a function
    of redshift.  The mean $(g-r)$ colour of all quasars with a
    $W4$-detection is given by the blue line. As in
    Figure~\ref{fig:ERQs_magredshift}, the three groups are indicated
    separately: Type 1s (dark blue circles), Type 2s (purple) and EREW
    objects (dark green). The extremely red quasars are generally redder
    than the mean $W4$ detected population in $(g-r)$ colour. However, the
    EREW objects are the exception; they {\it are bluer} than the full $W4$ population at all
    redshifts that they are found. This suggests that we are perhaps not
    sensitive to all EREWs, but just those objects with redshifts such
    that there are a number of very strong lines in the $g$-band.
    
    Fig.~\ref{fig:ERQs_rW1234_colors_vs_redshift} presents the optical
    to infrared colours, $(r-W1)$, $(r-W2)$, $(r-W3)$ and $(r-W4)$, for
    the full $W4$ matched sample as a function of redshift. The quasars
    detected at 22$\mu$m generally have a $r-W1$ colour of 5 or more for
    redshifts $z=0-3$, and again are redder than the general population of
    quasars. However, some of the Type 2 objects do have $(r-W1)$ and
    $(r-W2)$ colours consistent with the general $W4$ detected
    population. However, all the extremely red quasars are considerably
    redder in $(r-W3)$ which samples the ratio of rest-frame UV/blue to
    $\sim 6-7 \mu$m light at redshift $z\sim 0.8$ and rest-frame far UV to
    $\sim 3.4 \mu$m colour at redshift $z \sim 2.5$. By design, our sample
    is much redder than the general $W4$ detected quasar population in
    $(r-W4)$ colour, the latter of which has $10 \lesssim (r-W4) \lesssim
    12$ for redshifts $z=0-3$.

    \begin{figure}
      \includegraphics[width=8.60cm, height=8.60cm, trim=1.0cm 0.9cm -0.4cm 1.2cm]
      {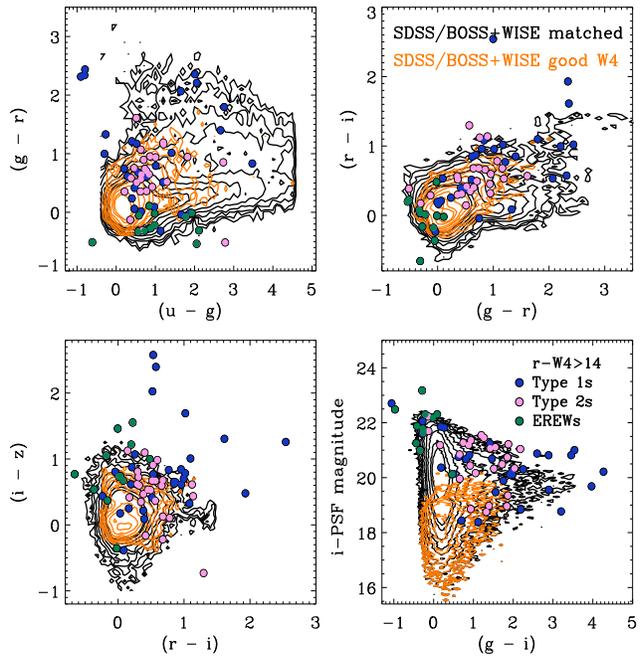}
      \centering
      \vspace{-14pt}
      \caption[]{
        The optical colour-colour properties of SDSS/BOSS+WISE matched
        sample and the extremely red quasars: 
        {\it (top left):}         $(u-g)$ vs. $(g-r)$; 
        {\it (top right):}         $(g-r)$ vs. $(r-i)$; 
        {\it (bottom left):}         $(r-i)$ vs. $(i-z)$; 
        {\it (bottom right):}         $(g-i)$ vs. $i$-band PSF magnitude. 
        The distribution of all the SDSS/BOSS quasars with a good match in any WISE band is given by the black contours, while those objects with good $W4$-matches, and their colour distributions are given by the solid orange contours. We also show the location in colour-colour space of the three broad groups of our extremely red quasars -- Type 1s (dark blue circles), Type 2s (purple) and EREW objects (dark green). The EREWs occupy a particular place in $gri$-colour-colour space, being very blue in $(g-i)$ colour.}
      \label{fig:colorcolor_ugriz}
    \end{figure}

    Figure~\ref{fig:colorcolor_ugriz} shows the optical colour-colour
    distribution of the full SDSS/BOSS WISE matched sample (black
    contours), and those that have good $W4$ detections (orange
    contours). The extremely red quasar population tends to avoid the
    $(u-g)\leqslant0.4$ and $(g-r)\leqslant0.3$ region. The EREWs occupy a
    particular location in $gri$ colour-colour and $(g-i)$-$i$
    colour-magnitude space. The EREWs have a very blue colour of
    $(g-i)\sim0$, in contrast to the overall extremely red quasar
    population, which tend to have redder $(g-i)$ colours than the general
    $W4$-matched sample. Further investigations into the colours and
    selection properties of the extremely red quasars and the EREWs are
    presented in Hamann et al. (in prep.).
    
    Note that $z\approx1.5$ is the redshift where \civ begins to
    contribute to the flux in the SDSS $g$-band. Thus, Type 2 quasars at
    $z\lesssim1.5$ may have colours inconsistent with the SDSS/BOSS
    optical quasar selection criteria and thus enter the SDSS/BOSS sample
    via their radio properties. On the other hand, Type 2 quasars at
    $z\gtrsim1.5$ with \civ and \lya emission really have blue continua.

\section{Summary and Conclusions}
\label{sec:conclusions}
We have matched the quasar catalogues of the SDSS and BOSS with WISE to
identify quasars with extremely high infrared-to-optical ratios
$r_{\rm AB}-W4_{\rm Vega}>14$ mag (i.e., $F_\nu({\rm 22\mu
m})/F_\nu(r) \simgt 1000$). We identify 65 objects and 
note the following findings and conclusions:

\begin{itemize}
    \item{This sample spans a redshift range of $0.28 < z < 4.36$ and has a bimodal distribution, with peaks 
        at $z\sim0.8$ and $z\sim2.5$.}
   \item{We recover a wide range of quasar spectra in this selection.  
        The majority of the objects have spectra of reddened Type 1
        quasars, Type 2 quasars (both at low and high redshift) and objects
        with strong absorption features.} 
    \item{There is a relatively high fraction of Type 2 objects at low redshift,
        suggesting that a high optical-to-infrared colour can be an efficient
        selection of narrow-line quasars.}
    \item{There are three objects that are detected in the $W4$-band but
        not $W1$ or $W2$ (i.e., ``W1W2-dropouts''), all of which are at
        $z>2.6$.}
    \item{We identify an intriguing class of objects at $z\simeq 2-3$ which are
        characterized by equivalent widths of REW(C\,{\sc iv})
        $\gtrsim150$\AA.  These objects often also have unusual line
        properties.  We speculate that the large REWs may be caused by
        suppressed continuum emission analogous to Type 2 quasars in the
        Unified Model. However, there is no obvious mechanism in the Unified
        Model to suppress the continuum without also suppressing the broad
        emission lines, thus potentially providing an interesting challenge to
        quasar models.} 
\end{itemize}

There are numerous avenues that the current dataset can contribute to
understanding quasar--host galaxy evolution and quasar orientation.
Hamann et al. (in prep.) investigate the nature and origins of extreme
broad emission line REWs in BOSS quasars. This will involve the
detailed examination of emission line properties for a subset of
extreme REW quasars to gain a better understanding of the physical
conditions and possible transient nature of this phenomenon. With the
addition of $\sim$18,000 spectra of 22$\mu$m detected quasars, one can
significantly update \citet{Roseboom13} where those authors
investigate the range of covering factors, determined from the ratio
of IR to UV/optical luminosity, seen in luminous Type 1 quasars.
However, the inhomogeneous selection of our dataset makes computing
the completeness challenging.  One can also measure the clustering of
the red-selected quasars, \citep[e.g.,][]{Donoso14, DiPompeo14}, using
spectroscopic redshifts to enable a more robust 3D-clustering
measurement. These investigations into enviromental, and host galaxy,
e.g. morphology, star-formation, properties of the very red
optical--IR colour-selected population will be able to elucidate on
the nature of the ``extremely red quasar'' population.

\section*{Acknowledgements}
We thank R. M. Cutri and the IPAC team for the Explanatory Supplement to the WISE All-Sky and AllWISE Data Release Products resource. Peregrine McGehee and the IRSA HelpDesk were also very useful.
N.P.R. thanks Nathan Bourne for useful discussions on infrared-to-radio flux ratios. 
AllWISE makes use of data from WISE, which is a joint project of the University of California, Los Angeles, and the Jet Propulsion Laboratory/California Institute of Technology, and NEOWISE, which is a project of the Jet Propulsion Laboratory/California Institute of Technology. WISE and NEOWISE are funded by the National Aeronautics and Space Administration.
N.P.R acknowledges that funding for part of this project was supplied by NASA and the {\it
Hubble} Grant Number: HST-GO-13014.06.
FH acknowledges support from the USA National Science Foundation grant AST-1009628.  

Funding for SDSS-III has been provided by the Alfred P. Sloan
Foundation, the Participating Institutions, the National Science
Foundation, and the U.S. Department of Energy. The SDSS-III web site
is \href{http://www.sdss3.org/}{http://www.sdss3.org/}.  SDSS-III is
managed by the Astrophysical Research Consortium for the Participating
Institutions of the SDSS-III Collaboration including the University of
Arizona, the Brazilian Participation Group, Brookhaven National
Laboratory, University of Cambridge, University of Florida, the French
Participation Group, the German Participation Group, the Instituto de
Astrofisica de Canarias, the Michigan State/Notre Dame/JINA
Participation Group, Johns Hopkins University, Lawrence Berkeley
National Laboratory, Max Planck Institute for Astrophysics, New Mexico
State University, New York University, Ohio State University,
Pennsylvania State University, University of Portsmouth, Princeton
University, the Spanish Participation Group, University of Tokyo,
University of Utah, Vanderbilt University, University of Virginia,
University of Washington, and Yale University.  \\ {\it Facilities:
SDSS, WISE}

\appendix 
\section{Notes on Individual Objects}

    \begin{figure}  
      \includegraphics[width=9.0cm, height=5.0cm, trim=28mm 00mm 00mm 5mm, clip]
      {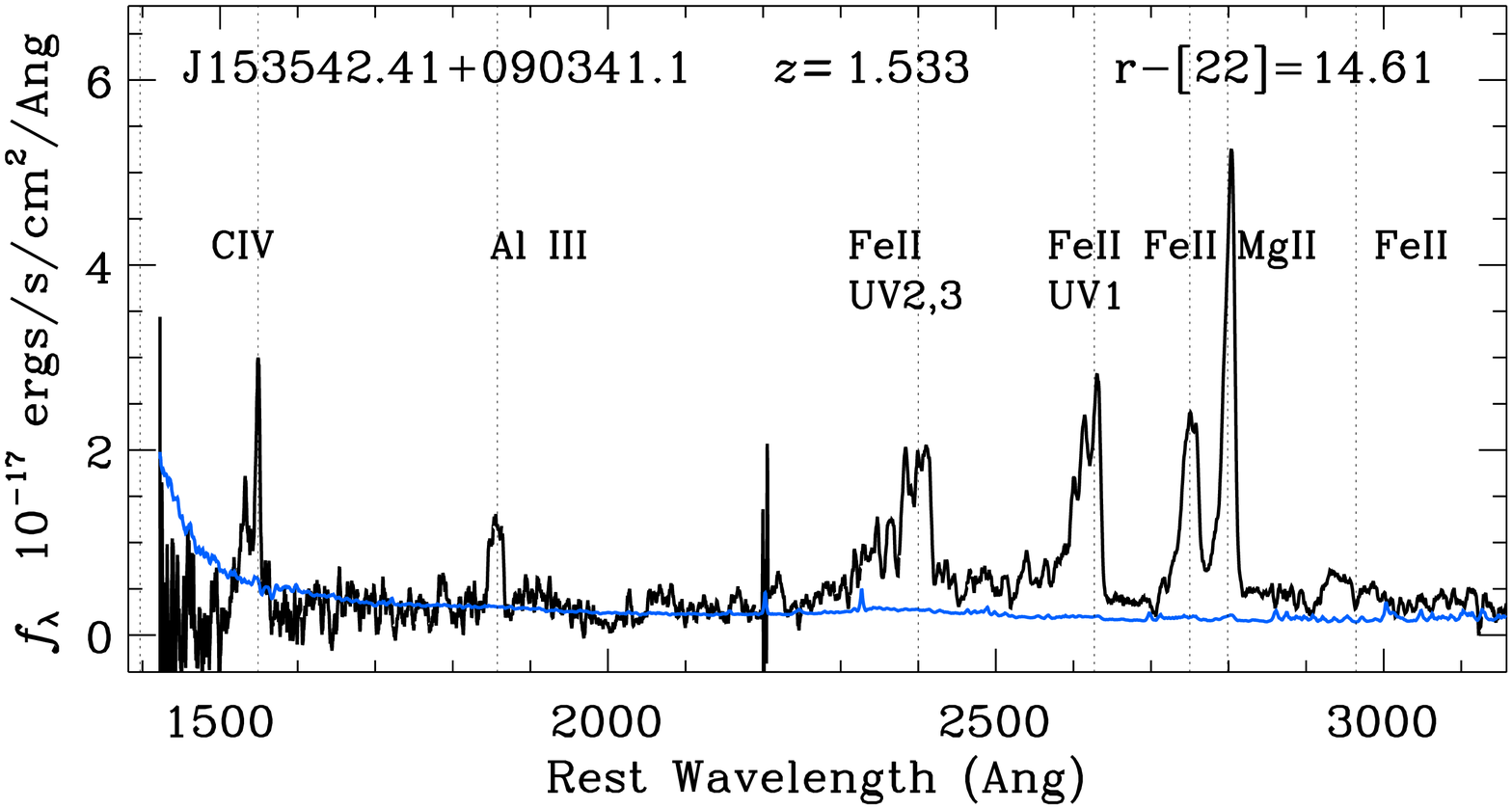}
      \centering
      \vspace{-8pt}
      \caption[]{The spectrum of SDSS J153542.41+090341.1. This object has
        exceptionally strong FeII emission in the UV1 (2600\AA), UV2/3
        (2400\AA) and 2740\AA\ complexes with weak emission from other FeII
        features.}
      \label{fig:J1535} 
    \end{figure}
    \subsection{J153542.41+090341.1}\label{sec:J1535}
    Figure~\ref{fig:J1535} displays the spectrum of SDSS
    J153542.41+090341.1, hereafter J1535+0903, a remarkable object with a
    unique pattern of emission lines. We classify this quasar as a Type 2
    candidate based on line widths in both \civ (Table 2) and \mgii
    (FWHM=$1600\pm60 \kms$). However, the exceptionally strong emission in
    \aliii 1860 and in \feii at $\sim2300-2770$\AA\ point to high
    densities indicative of a broad line region and thus a Type 1
    classification. In particular, photoionization models by
    \citet{Baldwin96} demonstrate that large ratios of \aliii 1860
    compared to the inter-combination lines with similar ionizations,
    \ciii 1909 and especially \siiii 1892, require densities $n_e \ga
    10^{12}$ cm$^{-3}$. Similarly, \citet{Baldwin04} showed that this
    peculiar pattern of FeII emission lines, dominated by a few spikes
    that include the resonance multiplets UV1 and UV 2,3, also requires
    densities $\ga$$10^{12}$ cm$^{-3}$ in a gas with very low ionization
    parameter. Hamann et al. (in prep.) discuss this object further. 

    \subsection{J013435.66-093102.9}\label{sec:J0134} 
    SDSS J013435.66-093102.9, hereafter J0134-0931, is a gravitational lens. J0134-0931 has $r_{\rm psf}=21.33$, $W4=6.59$ and a redshift $z=2.22$ and is also detected in FIRST at 0.96 Jy (integrated flux). J013435 has previously been     reported and studied in detail in \citet{Gregg02, Hall02lens, Keeton01, Keeton03, Winn02} and \citet{Winn03}.
    The flux is the linear sum of the lensing galaxy and the (boosted) quasar
    and at least one of the galaxy or quasar is exceedingly red. If there
    is dust in the foreground galaxy, it will extinct the quasar, and
    cause the $r-W4$ colour to be red. However, the foreground galaxy,
    at $z=0.76$, is faint enough to not contribute significantly to the
    overall flux, so the background quasar is extremely red on its own.


\bibliographystyle{mn2e}
\bibliography{tester_mnras}

\begin{thebibliography}{}

\bibitem[\protect\citeauthoryear{{Abazajian} et~al.,}{{Abazajian}
  et~al.}{2009}]{DR7}
{Abazajian} K.~N.,  et~al., 2009, \apjs, 182, 543

\bibitem[\protect\citeauthoryear{{Adelman-McCarthy} et~al.,}{{Adelman-McCarthy}
   et~al.}{2006}]{DR4}
{Adelman-McCarthy} J.~K.,  et~al., 2006, \apjs, 162, 38

\bibitem[\protect\citeauthoryear{{Ahn} et~al.,}{{Ahn}  et~al.}{2012}]{DR9}
{Ahn} C.~P.,  et~al., 2012, \apjs, 203, 21

\bibitem[\protect\citeauthoryear{{Ahn} et~al.,}{{Ahn}  et~al.}{2014}]{DR10}
{Ahn} C.~P.,  et~al., 2014, \apjs, 211, 17

\bibitem[\protect\citeauthoryear{{Alam} et~al.,}{{Alam}  et~al.}{2015}]{DR12}
{Alam} S.,  et~al., 2015, \apjs, 219, 12

\bibitem[\protect\citeauthoryear{{Alexandroff} et~al.,}{{Alexandroff}
  et~al.}{2013}]{Alexandroff13}
{Alexandroff} R.,  et~al., 2013, \mnras, 435, 3306

\bibitem[\protect\citeauthoryear{{Allen}, {Hewett}, {Maddox}, {Richards} \&
  {Belokurov}}{{Allen} et~al.}{2011}]{Allen11}
{Allen} J.~T.,  {Hewett} P.~C.,  {Maddox} N.,  {Richards} G.~T.,    {Belokurov}
  V.,  2011, \mnras, 410, 860

\bibitem[\protect\citeauthoryear{{Antonucci}}{{Antonucci}}{1993}]{Antonucci93}
{Antonucci} R.,  1993, \araa, 31, 473

\bibitem[\protect\citeauthoryear{{Antonucci} \& {Miller}}{{Antonucci} \&
  {Miller}}{1985}]{Antonucci_Miller85}
{Antonucci} R.~R.~J.,  {Miller} J.~S.,  1985, \apj, 297, 621

\bibitem[\protect\citeauthoryear{{Assef} et~al.,}{{Assef}
  et~al.}{2013}]{Assef13}
{Assef} R.~J.,  et~al., 2013, \apj, 772, 26

\bibitem[\protect\citeauthoryear{{Baldwin}, {Ferland}, {Korista}, {Carswell},
  {Hamann}, {Phillips}, {Verner}, {Wilkes} \& {Williams}}{{Baldwin}
  et~al.}{1996}]{Baldwin96}
{Baldwin} J.~A.,  {Ferland} G.~J.,  {Korista} K.~T.,  {Carswell} R.~F.,
  {Hamann} F.,  {Phillips} M.~M.,  {Verner} D.,  {Wilkes} B.~J.,    {Williams}
  R.~E.,  1996, \apj, 461, 664

\bibitem[\protect\citeauthoryear{{Baldwin}, {Ferland}, {Korista}, {Hamann} \&
  {LaCluyz{\'e}}}{{Baldwin} et~al.}{2004}]{Baldwin04}
{Baldwin} J.~A.,  {Ferland} G.~J.,  {Korista} K.~T.,  {Hamann} F.,
  {LaCluyz{\'e}} A.,  2004, \apj, 615, 610

\bibitem[\protect\citeauthoryear{{Baldwin}, {Phillips} \&
  {Terlevich}}{{Baldwin} et~al.}{1981}]{BPT}
{Baldwin} J.~A.,  {Phillips} M.~M.,    {Terlevich} R.,  1981, \pasp, 93, 5

\bibitem[\protect\citeauthoryear{{Banerji}, {Fabian} \& {McMahon}}{{Banerji}
  et~al.}{2014}]{Banerji14}
{Banerji} M.,  {Fabian} A.~C.,    {McMahon} R.~G.,  2014, \mnras

\bibitem[\protect\citeauthoryear{{Banerji}, {McMahon}, {Hewett},
  {Gonzalez-Solares} \& {Koposov}}{{Banerji} et~al.}{2013}]{Banerji13}
{Banerji} M.,  {McMahon} R.~G.,  {Hewett} P.~C.,  {Gonzalez-Solares} E.,
  {Koposov} S.~E.,  2013, \mnras, 429, L55

\bibitem[\protect\citeauthoryear{{Becker}, {White} \& {Helfand}}{{Becker}
  et~al.}{1995}]{Becker95}
{Becker} R.~H.,  {White} R.~L.,    {Helfand} D.~J.,  1995, \apj, 450, 559

\bibitem[\protect\citeauthoryear{{Bell}}{{Bell}}{2003}]{Bell03_SFRs}
{Bell} E.~F.,  2003, \apj, 586, 794

\bibitem[\protect\citeauthoryear{{Bolton} et~al.,}{{Bolton}
  et~al.}{2012}]{Bolton12}
{Bolton} A.~S.,  et~al., 2012, \aj, 144, 144

\bibitem[\protect\citeauthoryear{{Brand} et~al.,}{{Brand}
  et~al.}{2006}]{Brand06}
{Brand} K.,  et~al., 2006, \apj, 644, 143

\bibitem[\protect\citeauthoryear{{Brand} et~al.,}{{Brand}
  et~al.}{2007}]{Brand07}
{Brand} K.,  et~al., 2007, \apj, 663, 204

\bibitem[\protect\citeauthoryear{{Brand} et~al.,}{{Brand}
  et~al.}{2008}]{Brand08}
{Brand} K.,  et~al., 2008, \apj, 680, 119

\bibitem[\protect\citeauthoryear{{Brandt} \& {Hasinger}}{{Brandt} \&
  {Hasinger}}{2005}]{Brandt_Hasinger05}
{Brandt} W.~N.,  {Hasinger} G.,  2005, \araa, 43, 827

\bibitem[\protect\citeauthoryear{{Brotherton} et~al.,}{{Brotherton}
  et~al.}{1999}]{Brotherton99}
{Brotherton} M.~S.,  et~al., 1999, \apjl, 520, L87

\bibitem[\protect\citeauthoryear{{Brotherton}, {Grabelsky}, {Canalizo}, {van
  Breugel}, {Filippenko}, {Croom}, {Boyle} \& {Shanks}}{{Brotherton}
  et~al.}{2002}]{Brotherton02}
{Brotherton} M.~S.,  {Grabelsky} M.,  {Canalizo} G.,  {van Breugel} W.,
  {Filippenko} A.~V.,  {Croom} S.,  {Boyle} B.,    {Shanks} T.,  2002, \pasp,
  114, 593

\bibitem[\protect\citeauthoryear{{Bussmann} et~al.,}{{Bussmann}
  et~al.}{2009a}]{Bussmann09}
{Bussmann} R.~S.,  et~al., 2009a, \apj, 693, 750

\bibitem[\protect\citeauthoryear{{Bussmann} et~al.,}{{Bussmann}
  et~al.}{2009b}]{Bussmann09b}
{Bussmann} R.~S.,  et~al., 2009b, \apj, 705, 184

\bibitem[\protect\citeauthoryear{{Cales}, {Brotherton}, {Shang}, {Bennert},
  {Canalizo} \& {Diamond-Stanic}}{{Cales} et~al.}{2014}]{Cales14AAS}
{Cales} S.,  {Brotherton} M.~S.,  {Shang} Z.,  {Bennert} V.,  {Canalizo} G.,
  {Diamond-Stanic} A.~M.,  2014 Vol.~223 of AAS, \#115.06

\bibitem[\protect\citeauthoryear{{Cales} et~al.,}{{Cales}
  et~al.}{2013}]{Cales13}
{Cales} S.~L.,  et~al., 2013, \apj, 762, 90

\bibitem[\protect\citeauthoryear{{Casey}, {Narayanan} \& {Cooray}}{{Casey}
  et~al.}{2014}]{Casey14}
{Casey} C.~M.,  {Narayanan} D.,    {Cooray} A.,  2014, \physrep, 541, 45

\bibitem[\protect\citeauthoryear{{Chiu}, {Richards}, {Hewett} \&
  {Maddox}}{{Chiu} et~al.}{2007}]{Chiu07}
{Chiu} K.,  {Richards} G.~T.,  {Hewett} P.~C.,    {Maddox} N.,  2007, \mnras,
  375, 1180

\bibitem[\protect\citeauthoryear{{Cutri} et~al.,}{{Cutri}
  et~al.}{2011}]{Cutri11}
{Cutri} R.~M.,  et~al., 2011, Technical report, {Explanatory Supplement to the
  WISE Preliminary Data Release Products}

\bibitem[\protect\citeauthoryear{{Dawson} et~al.,}{{Dawson}
  et~al.}{2013}]{Dawson13}
{Dawson} K.,  et~al., 2013, \aj, 145, 10

\bibitem[\protect\citeauthoryear{{Dey} et~al.,}{{Dey}  et~al.}{2008}]{Dey08}
{Dey} A.,  et~al., 2008, \apj, 677, 943

\bibitem[\protect\citeauthoryear{{Diamond-Stanic} \& {Rieke}}{{Diamond-Stanic}
  \& {Rieke}}{2010}]{Diamond-Stanic10}
{Diamond-Stanic} A.~M.,  {Rieke} G.~H.,  2010, \apj, 724, 140

\bibitem[\protect\citeauthoryear{{DiPompeo}, {Myers}, {Hickox}, {Geach} \&
  {Hainline}}{{DiPompeo} et~al.}{2014}]{DiPompeo14}
{DiPompeo} M.~A.,  {Myers} A.~D.,  {Hickox} R.~C.,  {Geach} J.~E.,
  {Hainline} K.~N.,  2014, \mnras, 442, 3443

\bibitem[\protect\citeauthoryear{{Donley} et~al.,}{{Donley}
  et~al.}{2012}]{Donley12}
{Donley} J.~L.,  et~al., 2012, \apj, 748, 142

\bibitem[\protect\citeauthoryear{{Donoso}, {Yan}, {Stern} \& {Assef}}{{Donoso}
  et~al.}{2014}]{Donoso14}
{Donoso} E.,  {Yan} L.,  {Stern} D.,    {Assef} R.~J.,  2014, \apj, 789, 44

\bibitem[\protect\citeauthoryear{{Draine} \& {Li}}{{Draine} \&
  {Li}}{2007}]{Draine_Li07}
{Draine} B.~T.,  {Li} A.,  2007, \apj, 657, 810

\bibitem[\protect\citeauthoryear{{Eisenhardt} et~al.,}{{Eisenhardt}
  et~al.}{2012}]{Eisenhardt12}
{Eisenhardt} P.~R.~M.,  et~al., 2012, \apj, 755, 173

\bibitem[\protect\citeauthoryear{{Eisenstein}, {Weinberg} et~al.,}{{Eisenstein}
  et~al.}{2011}]{Eisenstein11}
{Eisenstein} D.~J.,  {Weinberg} D.~H.,    et~al., 2011, \aj, 142, 72

\bibitem[\protect\citeauthoryear{{Elitzur}}{{Elitzur}}{2008}]{Elitzur08}
{Elitzur} M.,  2008, \nar, 52, 274

\bibitem[\protect\citeauthoryear{{Elvis}}{{Elvis}}{2010}]{Elvis10}
{Elvis} M.,  2010, in IAU Symposium Vol.~267, {The Quasar Continuum}.
pp 55--64

\bibitem[\protect\citeauthoryear{{Elvis} et~al.,}{{Elvis}
  et~al.}{1994}]{Elvis94}
{Elvis} M.,  et~al., 1994, \apjs, 95, 1

\bibitem[\protect\citeauthoryear{{Fu} et~al.,}{{Fu}  et~al.}{2013}]{Fu13}
{Fu} H.,  et~al., 2013, \nat, 498, 338

\bibitem[\protect\citeauthoryear{{Fukugita}, {Ichikawa}, {Gunn}, {Doi},
  {Shimasaku} \& {Schneider}}{{Fukugita} et~al.}{1996}]{Fukugita96}
{Fukugita} M.,  {Ichikawa} T.,  {Gunn} J.~E.,  {Doi} M.,  {Shimasaku} K.,
  {Schneider} D.~P.,  1996, \aj, 111, 1748

\bibitem[\protect\citeauthoryear{{Fynbo}, {Krogager}, {Venemans}, {Noterdaeme},
  {Vestergaard}, {M{\o}ller}, {Ledoux} \& {Geier}}{{Fynbo}
  et~al.}{2013}]{Fynbo13}
{Fynbo} J.~P.~U.,  {Krogager} J.-K.,  {Venemans} B.,  {Noterdaeme} P.,
  {Vestergaard} M.,  {M{\o}ller} P.,  {Ledoux} C.,    {Geier} S.,  2013, \apjs,
  204, 6

\bibitem[\protect\citeauthoryear{{Gibson} et~al.,}{{Gibson}
  et~al.}{2009}]{Gibson09}
{Gibson} R.~R.,  et~al., 2009, \apj, 692, 758

\bibitem[\protect\citeauthoryear{{Glikman} et~al.,}{{Glikman}
  et~al.}{2012}]{Glikman12}
{Glikman} E.,  et~al., 2012, \apj, 757, 51

\bibitem[\protect\citeauthoryear{{Glikman} et~al.,}{{Glikman}
  et~al.}{2013}]{Glikman13}
{Glikman} E.,  et~al., 2013, \apj, 778, 127

\bibitem[\protect\citeauthoryear{{Glikman}, {Gregg}, {Lacy}, {Helfand},
  {Becker} \& {White}}{{Glikman} et~al.}{2004}]{Glikman04}
{Glikman} E.,  {Gregg} M.~D.,  {Lacy} M.,  {Helfand} D.~J.,  {Becker} R.~H.,
  {White} R.~L.,  2004, \apj, 607, 60

\bibitem[\protect\citeauthoryear{{Glikman}, {Helfand}, {White}, {Becker},
  {Gregg} \& {Lacy}}{{Glikman} et~al.}{2007}]{Glikman07}
{Glikman} E.,  {Helfand} D.~J.,  {White} R.~L.,  {Becker} R.~H.,  {Gregg}
  M.~D.,    {Lacy} M.,  2007, \apj, 667, 673

\bibitem[\protect\citeauthoryear{{Gordon} \& {Clayton}}{{Gordon} \&
  {Clayton}}{1998}]{Gordon98}
{Gordon} K.~D.,  {Clayton} G.~C.,  1998, \apj, 500, 816

\bibitem[\protect\citeauthoryear{{Greene} et~al.,}{{Greene}
  et~al.}{2014}]{Greene14}
{Greene} J.~E.,  et~al., 2014, \apj, 788, 91

\bibitem[\protect\citeauthoryear{{Gregg}, {Lacy}, {White}, {Glikman},
  {Helfand}, {Becker} \& {Brotherton}}{{Gregg} et~al.}{2002}]{Gregg02}
{Gregg} M.~D.,  {Lacy} M.,  {White} R.~L.,  {Glikman} E.,  {Helfand} D.,
  {Becker} R.~H.,    {Brotherton} M.~S.,  2002, \apj, 564, 133

\bibitem[\protect\citeauthoryear{{Groves}, {Heckman} \& {Kauffmann}}{{Groves}
  et~al.}{2006}]{Groves06}
{Groves} B.~A.,  {Heckman} T.~M.,    {Kauffmann} G.,  2006, \mnras, 371, 1559

\bibitem[\protect\citeauthoryear{{Gunn} et~al.,}{{Gunn}  et~al.}{1998}]{Gunn98}
{Gunn} J.~E.,  et~al., 1998, \aj, 116, 3040

\bibitem[\protect\citeauthoryear{{Gunn} et~al.,}{{Gunn}  et~al.}{2006}]{Gunn06}
{Gunn} J.~E.,  et~al., 2006, \aj, 131, 2332

\bibitem[\protect\citeauthoryear{{Hall} et~al.,}{{Hall}
  et~al.}{2002a}]{Hall02lens}
{Hall} P.~B.,  et~al., 2002a, \apjl, 575, L51

\bibitem[\protect\citeauthoryear{{Hall} et~al.,}{{Hall}
  et~al.}{2002b}]{Hall02}
{Hall} P.~B.,  et~al., 2002b, \apjs, 141, 267

\bibitem[\protect\citeauthoryear{{Hall}, {Martini}, {Depoy} \& {Gatley}}{{Hall}
  et~al.}{1997}]{Hall97}
{Hall} P.~B.,  {Martini} P.,  {Depoy} D.~L.,    {Gatley} I.,  1997, \apjl, 484,
  L17

\bibitem[\protect\citeauthoryear{{Harrison}, {Alexander}, {Mullaney} \&
  {Swinbank}}{{Harrison} et~al.}{2014}]{Harrison14}
{Harrison} C.~M.,  {Alexander} D.~M.,  {Mullaney} J.~R.,    {Swinbank} A.~M.,
  2014, \mnras, 441, 3306

\bibitem[\protect\citeauthoryear{{Ho}}{{Ho}}{2005}]{Ho05}
{Ho} L.~C.,  2005, \apj, 629, 680

\bibitem[\protect\citeauthoryear{{Hopkins} et~al.,}{{Hopkins}
  et~al.}{2005}]{Hopkins05b}
{Hopkins} P.~F.,  et~al., 2005, \apj, 630, 705

\bibitem[\protect\citeauthoryear{{Hopkins} et~al.,}{{Hopkins}
  et~al.}{2006}]{Hopkins06a}
{Hopkins} P.~F.,  et~al., 2006, \apjs, 163, 1

\bibitem[\protect\citeauthoryear{{Ivezi{\'c}} et~al.,}{{Ivezi{\'c}}
  et~al.}{2002}]{Ivezic02}
{Ivezi{\'c}} {\v Z}.,  et~al., 2002, \aj, 124, 2364

\bibitem[\protect\citeauthoryear{{Jarrett} et~al.,}{{Jarrett}
  et~al.}{2011}]{Jarrett11}
{Jarrett} T.~H.,  et~al., 2011, \apj, 735, 112

\bibitem[\protect\citeauthoryear{{Jones} et~al.,}{{Jones}
  et~al.}{2014}]{Jones14}
{Jones} S.~F.,  et~al., 2014, \mnras, 443, 146

\bibitem[\protect\citeauthoryear{{Jones} et~al.,}{{Jones}
  et~al.}{2015}]{Jones15}
{Jones} S.~F.,  et~al., 2015, \mnras, 448, 3325

\bibitem[\protect\citeauthoryear{{Jurek}, {Drinkwater}, {Francis} \&
  {Pimbblet}}{{Jurek} et~al.}{2008}]{Jurek08}
{Jurek} R.~J.,  {Drinkwater} M.~J.,  {Francis} P.~J.,    {Pimbblet} K.~A.,
  2008, \mnras, 383, 673

\bibitem[\protect\citeauthoryear{{Kauffmann} et~al.,}{{Kauffmann}
  et~al.}{2003}]{Kauffmann03}
{Kauffmann} G.,  et~al., 2003, \mnras, 346, 1055

\bibitem[\protect\citeauthoryear{{Keeton}}{{Keeton}}{2001}]{Keeton01}
{Keeton} C.~R.,  2001, ArXiv Astrophysics e-prints

\bibitem[\protect\citeauthoryear{{Keeton} \& {Winn}}{{Keeton} \&
  {Winn}}{2003}]{Keeton03}
{Keeton} C.~R.,  {Winn} J.~N.,  2003, \apj, 590, 39

\bibitem[\protect\citeauthoryear{{Kellermann}, {Sramek}, {Schmidt}, {Shaffer}
  \& {Green}}{{Kellermann} et~al.}{1989}]{Kellermann89}
{Kellermann} K.~I.,  {Sramek} R.,  {Schmidt} M.,  {Shaffer} D.~B.,    {Green}
  R.,  1989, \aj, 98, 1195

\bibitem[\protect\citeauthoryear{{Kennicutt}
  Jr.}{{Kennicutt}}{1998}]{Kennicutt98}
{Kennicutt} Jr. R.~C.,  1998, \araa, 36, 189

\bibitem[\protect\citeauthoryear{{Kewley} \& {Dopita}}{{Kewley} \&
  {Dopita}}{2002}]{Kewley02}
{Kewley} L.~J.,  {Dopita} M.~A.,  2002, \apjs, 142, 35

\bibitem[\protect\citeauthoryear{{Kewley}, {Geller} \& {Jansen}}{{Kewley}
  et~al.}{2004}]{Kewley04}
{Kewley} L.~J.,  {Geller} M.~J.,    {Jansen} R.~A.,  2004, \aj, 127, 2002

\bibitem[\protect\citeauthoryear{{Khachikian} \& {Weedman}}{{Khachikian} \&
  {Weedman}}{1974}]{Khachikian_Weedman74}
{Khachikian} E.~Y.,  {Weedman} D.~W.,  1974, \apj, 192, 581

\bibitem[\protect\citeauthoryear{{Kirkpatrick} et~al.,}{{Kirkpatrick}
  et~al.}{2012}]{Kirkpatrick12}
{Kirkpatrick} A.,  et~al., 2012, \apj, 759, 139

\bibitem[\protect\citeauthoryear{{Kirkpatrick} et~al.,}{{Kirkpatrick}
  et~al.}{2011}]{Kirkpatrick_JDavy11}
{Kirkpatrick} J.~D.,  et~al., 2011, \apjs, 197, 19

\bibitem[\protect\citeauthoryear{{Kocevski} et~al.,}{{Kocevski}
  et~al.}{2012}]{Kocevski12}
{Kocevski} D.~D.,  et~al., 2012, \apj, 744, 148

\bibitem[\protect\citeauthoryear{{Kratzer} \& {Richards}}{{Kratzer} \&
  {Richards}}{2015}]{Kratzer15}
{Kratzer} R.~M.,  {Richards} G.~T.,  2015, \aj, 149, 61

\bibitem[\protect\citeauthoryear{{Krawczyk}, {Richards}, {Mehta}, {Vogeley},
  {Gallagher}, {Leighly}, {Ross} \& {Schneider}}{{Krawczyk}
  et~al.}{2013}]{Krawczyk13}
{Krawczyk} C.~M.,  {Richards} G.~T.,  {Mehta} S.~S.,  {Vogeley} M.~S.,
  {Gallagher} S.~C.,  {Leighly} K.~M.,  {Ross} N.~P.,    {Schneider} D.~P.,
  2013, \apjs, 206, 4

\bibitem[\protect\citeauthoryear{{Krawczyk}, {Richards}, {Mehta}, {Vogeley},
  {Gallagher}, {Leighly}, {Ross} \& {Schneider}}{{Krawczyk}
  et~al.}{2015}]{Krawczyk15}
{Krawczyk} C.~M.,  {Richards} G.~T.,  {Mehta} S.~S.,  {Vogeley} M.~S.,
  {Gallagher} S.~C.,  {Leighly} K.~M.,  {Ross} N.~P.,    {Schneider} D.~P.,
  2015, \apjs, 999, 4

\bibitem[\protect\citeauthoryear{{Lacy} et~al.,}{{Lacy}  et~al.}{2004}]{Lacy04}
{Lacy} M.,  et~al., 2004, \apjs, 154, 166

\bibitem[\protect\citeauthoryear{{Lawrence} et~al.,}{{Lawrence}
  et~al.}{2007}]{Lawrence07}
{Lawrence} A.,  et~al., 2007, \mnras, 379, 1599

\bibitem[\protect\citeauthoryear{{Lee} et~al.,}{{Lee}  et~al.}{2013}]{KGLee13}
{Lee} K.-G.,  et~al., 2013, \aj, 145, 69

\bibitem[\protect\citeauthoryear{{Liu}, {Zakamska}, {Greene}, {Nesvadba} \&
  {Liu}}{{Liu} et~al.}{2013}]{Liu13b}
{Liu} G.,  {Zakamska} N.~L.,  {Greene} J.~E.,  {Nesvadba} N.~P.~H.,    {Liu}
  X.,  2013, \mnras, 436, 2576

\bibitem[\protect\citeauthoryear{{Liu}, {Shen}, {Strauss} \& {Greene}}{{Liu}
  et~al.}{2010}]{Liu10}
{Liu} X.,  {Shen} Y.,  {Strauss} M.~A.,    {Greene} J.~E.,  2010, \apj, 708,
  427

\bibitem[\protect\citeauthoryear{{Maddox}, {Hewett}, {P{\'e}roux}, {Nestor} \&
  {Wisotzki}}{{Maddox} et~al.}{2012}]{Maddox12}
{Maddox} N.,  {Hewett} P.~C.,  {P{\'e}roux} C.,  {Nestor} D.~B.,    {Wisotzki}
  L.,  2012, \mnras, 424, 2876

\bibitem[\protect\citeauthoryear{{Maddox}, {Hewett}, {Warren} \&
  {Croom}}{{Maddox} et~al.}{2008}]{Maddox08}
{Maddox} N.,  {Hewett} P.~C.,  {Warren} S.~J.,    {Croom} S.~M.,  2008, \mnras,
  386, 1605

\bibitem[\protect\citeauthoryear{{Mainzer} et~al.,}{{Mainzer}
  et~al.}{2011}]{Mainzer11}
{Mainzer} A.,  et~al., 2011, \apj, 731, 55

\bibitem[\protect\citeauthoryear{{Marchesini} et~al.,}{{Marchesini}
  et~al.}{2007}]{Marchesini07}
{Marchesini} D.,  et~al., 2007, \apj, 656, 42

\bibitem[\protect\citeauthoryear{{Marchesini}, {Stefanon}, {Brammer} \&
  {Whitaker}}{{Marchesini} et~al.}{2012}]{Marchesini12}
{Marchesini} D.,  {Stefanon} M.,  {Brammer} G.~B.,    {Whitaker} K.~E.,  2012,
  \apj, 748, 126

\bibitem[\protect\citeauthoryear{{Mart{\'{\i}}nez-Sansigre}, {Rawlings},
  {Lacy}, {Fadda}, {Jarvis}, {Marleau}, {Simpson} \&
  {Willott}}{{Mart{\'{\i}}nez-Sansigre} et~al.}{2006}]{Martinez-Sansigre06}
{Mart{\'{\i}}nez-Sansigre} A.,  {Rawlings} S.,  {Lacy} M.,  {Fadda} D.,
  {Jarvis} M.~J.,  {Marleau} F.~R.,  {Simpson} C.,    {Willott} C.~J.,  2006,
  \mnras, 370, 1479

\bibitem[\protect\citeauthoryear{{Matsuoka} et~al.,}{{Matsuoka}
  et~al.}{2015}]{Matsuoka15}
{Matsuoka} Y.,  et~al., 2015, ArXiv e-prints

\bibitem[\protect\citeauthoryear{{Melbourne} et~al.,}{{Melbourne}
  et~al.}{2009}]{Melbourne09}
{Melbourne} J.,  et~al., 2009, \aj, 137, 4854

\bibitem[\protect\citeauthoryear{{Melbourne} et~al.,}{{Melbourne}
  et~al.}{2012}]{Melbourne12}
{Melbourne} J.,  et~al., 2012, \aj, 143, 125

\bibitem[\protect\citeauthoryear{{Mostek}, {Coil}, {Moustakas}, {Salim} \&
  {Weiner}}{{Mostek} et~al.}{2012}]{Mostek12}
{Mostek} N.,  {Coil} A.~L.,  {Moustakas} J.,  {Salim} S.,    {Weiner} B.~J.,
  2012, \apj, 746, 124

\bibitem[\protect\citeauthoryear{{Moustakas}, {Kennicutt} Jr. \&
  {Tremonti}}{{Moustakas} et~al.}{2006}]{Moustakas06}
{Moustakas} J.,  {Kennicutt} Jr. R.~C.,    {Tremonti} C.~A.,  2006, \apj, 642,
  775

\bibitem[\protect\citeauthoryear{{Mullaney}, {Alexander}, {Goulding} \&
  {Hickox}}{{Mullaney} et~al.}{2011}]{Mullaney11}
{Mullaney} J.~R.,  {Alexander} D.~M.,  {Goulding} A.~D.,    {Hickox} R.~C.,
  2011, \mnras, 414, 1082

\bibitem[\protect\citeauthoryear{{Nakos} et~al.,}{{Nakos}
  et~al.}{2009}]{Nakos09}
{Nakos} T.,  et~al., 2009, \aap, 494, 579

\bibitem[\protect\citeauthoryear{{Norman} et~al.,}{{Norman}
  et~al.}{2002}]{Norman02}
{Norman} C.,  et~al., 2002, \apj, 571, 218

\bibitem[\protect\citeauthoryear{{Oke} \& {Gunn}}{{Oke} \&
  {Gunn}}{1983}]{Oke83}
{Oke} J.~B.,  {Gunn} J.~E.,  1983, \apj, 266, 713

\bibitem[\protect\citeauthoryear{{Padovani}, {Perlman}, {Landt}, {Giommi} \&
  {Perri}}{{Padovani} et~al.}{2003}]{Padovani03}
{Padovani} P.,  {Perlman} E.~S.,  {Landt} H.,  {Giommi} P.,    {Perri} M.,
  2003, \apj, 588, 128

\bibitem[\protect\citeauthoryear{{P{\^a}ris} et~al.,}{{P{\^a}ris}
  et~al.}{2014}]{Paris14}
{P{\^a}ris} I.,  et~al., 2014, \aap, 563, A54

\bibitem[\protect\citeauthoryear{{Peth}, {Ross} \& {Schneider}}{{Peth}
  et~al.}{2011}]{Peth11}
{Peth} M.~A.,  {Ross} N.~P.,    {Schneider} D.~P.,  2011, \aj, 141, 105

\bibitem[\protect\citeauthoryear{{Polletta} et~al.,}{{Polletta}
  et~al.}{2008}]{Polletta08}
{Polletta} A.,  et~al., 2008, \apj, 675, 960

\bibitem[\protect\citeauthoryear{{Pope} et~al.,}{{Pope}  et~al.}{2008}]{Pope08}
{Pope} A.,  et~al., 2008, \apj, 689, 127

\bibitem[\protect\citeauthoryear{{Reyes} et~al.,}{{Reyes}
  et~al.}{2008}]{Reyes08}
{Reyes} R.,  et~al., 2008, \aj, 136, 2373

\bibitem[\protect\citeauthoryear{{Richards} et~al.,}{{Richards}
  et~al.}{2002}]{Richards02}
{Richards} G.~T.,  et~al., 2002, \aj, 123, 2945

\bibitem[\protect\citeauthoryear{{Richards} et~al.,}{{Richards}
  et~al.}{2003}]{Richards03}
{Richards} G.~T.,  et~al., 2003, \aj, 126, 1131

\bibitem[\protect\citeauthoryear{{Richards} et~al.,}{{Richards}
  et~al.}{2006}]{Richards06b}
{Richards} G.~T.,  et~al., 2006, \apjs, 166, 470

\bibitem[\protect\citeauthoryear{{Richards} et~al.,}{{Richards}
  et~al.}{2009}]{Richards09b}
{Richards} G.~T.,  et~al., 2009, \aj, 137, 3884

\bibitem[\protect\citeauthoryear{{Rosario} et~al.,}{{Rosario}
  et~al.}{2015}]{Rosario15}
{Rosario} D.~J.,  et~al., 2015, \aap, 573, A85

\bibitem[\protect\citeauthoryear{{Roseboom}, {Lawrence}, {Elvis}, {Petty},
  {Shen} \& {Hao}}{{Roseboom} et~al.}{2013}]{Roseboom13}
{Roseboom} I.~G.,  {Lawrence} A.,  {Elvis} M.,  {Petty} S.,  {Shen} Y.,
  {Hao} H.,  2013, \mnras, 429, 1494

\bibitem[\protect\citeauthoryear{{Ross} et~al.,}{{Ross}  et~al.}{2012}]{Ross12}
{Ross} N.~P.,  et~al., 2012, \apjs, 199, 3

\bibitem[\protect\citeauthoryear{{Sanders}, {Soifer}, {Elias}, {Madore},
  {Matthews}, {Neugebauer} \& {Scoville}}{{Sanders} et~al.}{1988}]{Sanders88}
{Sanders} D.~B.,  {Soifer} B.~T.,  {Elias} J.~H.,  {Madore} B.~F.,  {Matthews}
  K.,  {Neugebauer} G.,    {Scoville} N.~Z.,  1988, \apj, 325, 74

\bibitem[\protect\citeauthoryear{{Schlegel}, {Finkbeiner} \&
  {Davis}}{{Schlegel} et~al.}{1998}]{Schlegel98}
{Schlegel} D.~J.,  {Finkbeiner} D.~P.,    {Davis} M.,  1998, \apj, 500, 525

\bibitem[\protect\citeauthoryear{{Schneider} et~al.,}{{Schneider}
  et~al.}{2010}]{Schneider10}
{Schneider} D.~P.,  et~al., 2010, \aj, 139, 2360

\bibitem[\protect\citeauthoryear{{Shen} et~al.,}{{Shen}  et~al.}{2011}]{Shen11}
{Shen} Y.,  et~al., 2011, \apjs, 194, 45

\bibitem[\protect\citeauthoryear{{Smee} et~al.,}{{Smee}  et~al.}{2013}]{Smee13}
{Smee} S.~A.,  et~al., 2013, \aj, 146, 32

\bibitem[\protect\citeauthoryear{{Smith} et~al.,}{{Smith}
  et~al.}{2002}]{Smith02}
{Smith} J.~A.,  et~al., 2002, \aj, 123, 2121

\bibitem[\protect\citeauthoryear{{Soifer}, {Helou} \& {Werner}}{{Soifer}
  et~al.}{2008}]{Soifer08}
{Soifer} B.~T.,  {Helou} G.,    {Werner} M.,  2008, \araa, 46, 201

\bibitem[\protect\citeauthoryear{{Souchay} et~al.,}{{Souchay}
  et~al.}{2009}]{Souchay09}
{Souchay} J.,  et~al., 2009, \aap, 494, 799

\bibitem[\protect\citeauthoryear{{Stern} et~al.,}{{Stern}
  et~al.}{2002}]{Stern02}
{Stern} D.,  et~al., 2002, \apj, 568, 71

\bibitem[\protect\citeauthoryear{{Stern} et~al.,}{{Stern}
  et~al.}{2005}]{Stern05}
{Stern} D.,  et~al., 2005, \apj, 631, 163

\bibitem[\protect\citeauthoryear{{Stern} et~al.,}{{Stern}
  et~al.}{2012}]{Stern12}
{Stern} D.,  et~al., 2012, \apj, 753, 30

\bibitem[\protect\citeauthoryear{{Stoughton} et~al.,}{{Stoughton}
  et~al.}{2002}]{Stoughton02}
{Stoughton} C.,  et~al., 2002, \aj, 123, 485

\bibitem[\protect\citeauthoryear{{Trump} et~al.,}{{Trump}
  et~al.}{2006}]{Trump06}
{Trump} J.~R.,  et~al., 2006, \apjs, 165, 1

\bibitem[\protect\citeauthoryear{{Tsai} et~al.,}{{Tsai}  et~al.}{2015}]{Tsai15}
{Tsai} C.-W.,  et~al., 2015, \apj, 805, 90

\bibitem[\protect\citeauthoryear{{Urrutia}, {Lacy} \& {Becker}}{{Urrutia}
  et~al.}{2008}]{Urrutia08}
{Urrutia} T.,  {Lacy} M.,    {Becker} R.~H.,  2008, \apj, 674, 80

\bibitem[\protect\citeauthoryear{{Vanden Berk} et~al.,}{{Vanden Berk}
  et~al.}{2001}]{VdB01}
{Vanden Berk} D.~E.,  et~al., 2001, \aj, 122, 549

\bibitem[\protect\citeauthoryear{{Vardanyan}, {Weedman} \&
  {Sargsyan}}{{Vardanyan} et~al.}{2014}]{Vardanyan14}
{Vardanyan} V.,  {Weedman} D.,    {Sargsyan} L.,  2014, \apj, 790, 88

\bibitem[\protect\citeauthoryear{{Veilleux} et~al.,}{{Veilleux}
  et~al.}{2013}]{Veilleux13}
{Veilleux} S.,  et~al., 2013, \apj, 764, 15

\bibitem[\protect\citeauthoryear{{Veilleux} \& {Osterbrock}}{{Veilleux} \&
  {Osterbrock}}{1987}]{Veilleux87}
{Veilleux} S.,  {Osterbrock} D.~E.,  1987, \apjs, 63, 295

\bibitem[\protect\citeauthoryear{{Villforth} et~al.,}{{Villforth}
  et~al.}{2014}]{Villforth14}
{Villforth} C.,  et~al., 2014, \mnras, 439, 3342

\bibitem[\protect\citeauthoryear{{Wardlow} et~al.,}{{Wardlow}
  et~al.}{2013}]{Wardlow13}
{Wardlow} J.~L.,  et~al., 2013, \apj, 762, 59

\bibitem[\protect\citeauthoryear{{Wei}, {Shang}, {Brotherton}, {Cales},
  {Hines}, {Dale}, {Ganguly} \& {Canalizo}}{{Wei} et~al.}{2013}]{Wei13}
{Wei} P.,  {Shang} Z.,  {Brotherton} M.~S.,  {Cales} S.~L.,  {Hines} D.~C.,
  {Dale} D.~A.,  {Ganguly} R.,    {Canalizo} G.,  2013, \apj, 772, 28

\bibitem[\protect\citeauthoryear{{Weingartner} \& {Draine}}{{Weingartner} \&
  {Draine}}{2001}]{Weingartner01}
{Weingartner} J.~C.,  {Draine} B.~T.,  2001, \apj, 548, 296

\bibitem[\protect\citeauthoryear{{Weymann}, {Morris}, {Foltz} \&
  {Hewett}}{{Weymann} et~al.}{1991}]{Weymann91}
{Weymann} R.~J.,  {Morris} S.~L.,  {Foltz} C.~B.,    {Hewett} P.~C.,  1991,
  \apj, 373, 23

\bibitem[\protect\citeauthoryear{{White}, {Zheng}, {Brown}, {Dey} \&
  {Jannuzi}}{{White} et~al.}{2007}]{White07}
{White} M.,  {Zheng} Z.,  {Brown} M.~J.~I.,  {Dey} A.,    {Jannuzi} B.~T.,
  2007, \apjl, 655, L69

\bibitem[\protect\citeauthoryear{{White}, {Helfand}, {Becker}, {Glikman} \& {de
  Vries}}{{White} et~al.}{2007}]{White07radio}
{White} R.~L.,  {Helfand} D.~J.,  {Becker} R.~H.,  {Glikman} E.,    {de Vries}
  W.,  2007, \apj, 654, 99

\bibitem[\protect\citeauthoryear{{Winn}, {Kochanek}, {Keeton} \&
  {Lovell}}{{Winn} et~al.}{2003}]{Winn03}
{Winn} J.~N.,  {Kochanek} C.~S.,  {Keeton} C.~R.,    {Lovell} J.~E.~J.,  2003,
  \apj, 590, 26

\bibitem[\protect\citeauthoryear{{Winn}, {Lovell}, {Chen}, {Fletcher},
  {Hewitt}, {Patnaik} \& {Schechter}}{{Winn} et~al.}{2002}]{Winn02}
{Winn} J.~N.,  {Lovell} J.~E.~J.,  {Chen} H.-W.,  {Fletcher} A.~B.,  {Hewitt}
  J.~N.,  {Patnaik} A.~R.,    {Schechter} P.~L.,  2002, \apj, 564, 143

\bibitem[\protect\citeauthoryear{{Wright} et~al.,}{{Wright}
  et~al.}{2010}]{Wright10}
{Wright} E.~L.,  et~al., 2010, \aj, 140, 1868

\bibitem[\protect\citeauthoryear{{Wu} et~al.,}{{Wu}
  et~al.}{2012}]{Wu12_w1w2drops}
{Wu} J.,  et~al., 2012, \apj, 756, 96

\bibitem[\protect\citeauthoryear{{Wu} \& {Jia}}{{Wu} \& {Jia}}{2010}]{Wu10}
{Wu} X.,  {Jia} Z.,  2010, \mnras, 406, 1583

\bibitem[\protect\citeauthoryear{{Wu}, {Wang}, {Schmidt}, {Bian}, {Jiang} \&
  {Fan}}{{Wu} et~al.}{2011}]{Wu11}
{Wu} X.-B.,  {Wang} R.,  {Schmidt} K.~B.,  {Bian} F.,  {Jiang} L.,    {Fan} X.,
   2011, \aj, 142, 78

\bibitem[\protect\citeauthoryear{{Wu}, {Zuo}, {Yang}, {Yang} \& {Wang}}{{Wu}
  et~al.}{2013}]{Wu13}
{Wu} X.-B.,  {Zuo} W.,  {Yang} J.,  {Yang} Q.,    {Wang} F.,  2013, \aj, 146,
  100

\bibitem[\protect\citeauthoryear{{Yan} et~al.,}{{Yan}  et~al.}{2013}]{Yan13}
{Yan} L.,  et~al., 2013, \aj, 145, 55

\bibitem[\protect\citeauthoryear{{York} et~al.,}{{York}  et~al.}{2000}]{York00}
{York} D.~G.,  et~al., 2000, \aj, 120, 1579

\bibitem[\protect\citeauthoryear{{Zakamska} et~al.,}{{Zakamska}
  et~al.}{2003}]{Zakamska03}
{Zakamska} N.~L.,  et~al., 2003, \aj, 126, 2125

\bibitem[\protect\citeauthoryear{{Zakamska} et~al.,}{{Zakamska}
  et~al.}{2005}]{Zakamska05}
{Zakamska} N.~L.,  et~al., 2005, \aj, 129, 1212

\bibitem[\protect\citeauthoryear{{Zakamska} et~al.,}{{Zakamska}
  et~al.}{2006}]{Zakamska06}
{Zakamska} N.~L.,  et~al., 2006, \aj, 132, 1496

\bibitem[\protect\citeauthoryear{{Zakamska} et~al.,}{{Zakamska}
  et~al.}{2015}]{Zakamska15}
{Zakamska} N.~L.,  et~al., 2015

\bibitem[\protect\citeauthoryear{{Zakamska} \& {Greene}}{{Zakamska} \&
  {Greene}}{2014}]{Zakamska_Greene14}
{Zakamska} N.~L.,  {Greene} J.~E.,  2014, \mnras, 442, 784

\end{thebibliography}

\end{document}